\documentclass[oldversion]{aa} 
\usepackage{graphicx}
\usepackage{aalongtable}
\usepackage{txfonts}
\usepackage{rotating}
\usepackage{lscape}
\newcommand{\gsim}{\;\lower.6ex\hbox{$\sim$}\kern-7.75pt\raise.65ex\hbox{$>$}\;}
\newcommand{\lsim}{\;\lower.6ex\hbox{$\sim$}\kern-7.75pt\raise.65ex\hbox{$<$}\;}

\usepackage{amstext}

\begin{document}

\title{Excess of Ca (and Sc) produced in globular cluster multiple populations:
a first census in 77 Galactic globular clusters\thanks{Based on observations
collected at  ESO telescopes under programmes 072.D-0507, 073.D-0211,
081.D-0286, 083.D-0208, 087.B-0086, 093.B-0583, 095.B-0028, 099.D-0047
(proprietary); 073.D-0760,  381.D-0329,  095.D-0834, 095.D-0539  (archival).} }

\author{
Eugenio Carretta\inst{1}
\and
Angela Bragaglia\inst{1}
}

\authorrunning{Carretta and Bragaglia}
\titlerunning{Products from very hot H-burning}

\offprints{E. Carretta, eugenio.carretta@inaf.it}

\institute{
INAF-Osservatorio di Astrofisica e Scienza dello Spazio di Bologna, via Gobetti
 93/3, I-40129 Bologna, Italy}

\date{}

\abstract{Multiple stellar populations in globular clusters (GCs) are distinct
by their different abundances of light elements. The abundance anti-correlations
point towards a nucleosynthesis origin due to high-temperature H burning, but 
it remains to be assessed which type of stars altered primordial abundances in GCs. In particular,  the regime at very high temperature that shapes
the variations in potassium as well as calcium and scandium, which has been detected in a few
notable cases such as NGC~2419 and NGC~2808, is still poorly  explored. We
started a systematic search for excess of Ca (and Sc) in GC stars with respect
to the level of unmodified field stars. This method has recently been proven to be highly
efficient in revealing the outcome of the proton-capture reactions at very high
temperatures. Statistically robust evidence of such excess was found in a
small number of GCs (NGC~4833, NGC~6715, NGC~6402, NGC~5296, NGC~5824, and 
NGC~5139/$\omega$~Centauri) that join the previously known two clusters. For the
first time we show that NGC~4833 is likely to host anti-correlated K
and Mg abundances. All these GCs are among the most massive ones in the Galaxy. We
found that the fraction of stars with Ca enhancement at 3$\sigma$ above the
field star distribution is a multivariate function of the GC mass and
metallicity, as in other manifestations of the multiple population phenomenon in
GCs. We argue that these alterations in only a few GCs can be 
reproduced by two different channels: either a class of ordinary stars, that is common
to all GCs, acts only in particular environments, or an on-off mechanism is
generated by the occurrence of a peculiar type of stars (or lack of such stars). Hot
bottom-burning in asymptotic giant branch stars in the low-metallicity regime
is a good candidate for the first class. Alternatively, a metallicity dependence
is also expected for supermassive stars, which are predicted to preferentially
form in massive GCs.}

\maketitle

\section{Introduction}
The signature of multiple stellar populations in globular clusters (GCs) is the
star-to-star difference in the abundances of light elements (from C to Si, and
in a few cases, up to K, Ca, and Sc) in them and with respect to the field
stars of similar metallicity. The observational evidence accumulated in the past
half a century is reviewed in Gratton et al. (2004, 2012, 2019) and in Bastian
and Lardo  (2018). The whole chemical pattern of multiple populations and its
reflection  in the photometric sequences of the colour-magnitude diagrams
(CMD) can be explained as caused by the nucleosynthesis through
proton-capture reactions in H-burning at high temperature (Denisenkov and
Denisenkova 1989, Langer et al. 1993). 

This origin is strongly favoured because the alterations are not randomly
distributed in chemical space. The depletion in elements that are consumed
in proton-capture reactions (C, O, and Mg) is always accompanied by enhancement in
the abundance of species that  are produced by these same reactions (N, Na, Al,
and Si). This is obviously the reason why the main observed signature is represented
by the well-known anti-correlations C-N, O-Na, Mg-Al, etc., as well as
correlations between Na-Al, Al-Si, and so on. 
Some doubts about the nucleosynthetic origin of these patterns have been cast by
discrepancies with the output of theoretical models that tried to explain them.
However, the extreme regularity with which we observe O and Mg abundances that are only
depleted, never enhanced, or  viceversa, Na and Al only enhanced, never depleted
with respect to field halo stars, leaves us with no doubt that the mechanism
must reside in nucleosynthesis operating through proton-capture reactions. The
simple truth is that models still cannot faithfully reproduce the whole
complexity of GCs (e.g. Bastian et al. 2015).

A direct inference is that multiple stellar populations in GCs must also differ
in age, the amount of the difference depending on the proposed candidate
polluter for nuclearly processed matter. This stems from the inability of
currently evolving low-mass stars to achieve the high
temperatures for H-burning in their interior that are required to operate the proton-capture reactions, in
particular for the most energetic cycles, such as Mg-Al or the leakage
from this on $^{28}$Si (Karakas and Lattanzio 2003). This in turn implies that
the nucleosynthesis in question must have occurred in stars more
massive than the currently observed stars, which were born, evolved, and  ended their
life in previous stellar generations (e.g. Gratton et al. 2001). 

While the constraint on the temperature for nuclear burning is physically
robust, the  imperfect match between observations and models still precludes a
clear identification of the stellar sites at which this burning occurred. Proposed
candidate polluters cover a wide range in mass, ranging from  asymptotic giant
branch (AGB) stars (e.g. Cottrell and Da Costa 1981, Ventura et al. 2001)
to  massive fast-rotating stars (Decressin et al. 2007) or in binary
interaction (de Mink et al. 2009), and more exotic candidates such as very massive
stars (Denissenkov and Hartwick 2014, Gieles et al. 2018) and  supergiants
(Sz\'{e}csi and W\"unsch 2019).

Producing elements higher in mass than Si from proton-captures would require  very high
temperatures in order to overcome the Coulomb barrier. The interest in the group
K-Ca-Sc has constantly grown since the triggering discovery of large variations
in K abundances in NGC~2419. From high-resolution spectra,  Cohen and Kirby
(2012) found K, Ca, and Sc enhanced in Mg-poor stars of this cluster, while the
medium-resolution spectroscopic survey by Mucciarelli et al.  (2012) detected a
wide spread in Mg and K abundances, anti-correlated with each other, in giants
of NGC~2419 sharing the same Fe and Ca abundances.  Their analysis  revealed
that the star with huge [K/Fe] and high Mg depletion found in Cohen et al.
(2011) is not the only notable exception in this GC.

Initially,  the impact of K production in the context of multiple populations
was disregarded because it was thought to require temperatures too high outside supernovae.  
However, Ventura et al. (2012) advocated that the observed K-Mg anti-correlation
might be  explained by the simultaneous activation of the Mg-Al-Si and Ar-K
cycles in particular conditions, for example low metallicity and enhanced efficiency of
the hot bottom-burning in the AGB phase. A first scrutiny (Carretta et al.
2013a) of K abundances in a few other GCs, based on a limited number of stars in
each GC, appeared to confirm the uniqueness of the pattern in NGC~2419, as well
as the anti-correlation of Ca and Sc with Mg abundances in this GC (see also
Cohen and Kirby 2012). 

Mucciarelli et al. (2015) targeted NGC~2808, which is one of the GCs with the most
extended Na-O anti-correlation (Carretta et al. 2006), and found that K is not
only anti-correlated with Mg, but is also correlated with Al and Na abundances.
Together with the correlation Ca-Sc and the Sc-Mg anti-correlation discovered by
Carretta (2015) in NGC~2808, these observations left no doubt that K, Ca, and Sc
are affected by real variations linked to the proton-capture processes in
multiple population in GCs, even if their mass is lower than that of NGC~2419.

However, to fully explore this high-temperature regime at which the first-generation (FG) polluters operated in the early protoclusters, it is highly
desirable to gain more statistics.  Carretta et al. (2013a) have pointed out
that the paucity of K abundances available in GCs can be bypassed by exploiting
the Ca and Mg abundance determinations, which are available for many more stars and
are usually based on several atomic transitions. A first attempt was successfully
made in Carretta et al. (2014a), who plotted in their figure 10 the [Ca/Mg]
ratio as a function of [Ca/H] for more than 200 red giants in about 20 GCs.
In addition to NGC~2419, five GCs stand out in this plane, with stars showing an excess
of Ca with respect to Mg: NGC~4833, NGC~7078 (M~15), NGC~2808, NGC~6715 (M~54),
and NGC~5139 ($\omega$ Centauri).

With the same purpose, Carretta and Bragaglia (2019) defined a similar
diagnostic plot, this time based not on a comparison with GC stars, but using field stars. The idea was to obtain more efficient detection plots because  field halo stars are an almost pure FG population. Many recent studies
(e.g. Martell et al. 2011, Koch et al. 2019a) estimated the contribution of GC
stars with a second-generation (SG) composition only at the 2-2.5\% level in
the halo. The rarity of these stars makes it improbable that the control sample is
much affected by SG interlopers and thus can be safely
used as a baseline to probe the possible outcome of the multiple-population
phenomenon of the Ca and Sc anti-correlations with Mg. We tested the efficiency of
these diagnostics by  showing that a Ca excess is easily detected in the SG
stars of NGC~2808, as expected, but not in those of the massive bulge cluster
NGC~6388. In addition, we highlighted that this approach also provides an accurate enough estimate of the temperature range achievable by the FG polluters. When Al-Mg-Si variations are detected but no variations in Ca and Sc are
observed, the inferred temperatures must be high enough to activate the Mg-Al
cycle, possibly leaking on Si (T$>100-110$ MK), but not high enough to trigger the
K production from Ar (T$=120-150$ MK; Ventura et al. 2012, D'Antona et al.
2016,  Prantzos et al. 2017). In this way, we can hope to supply useful
additional  constraints to theoretical models to devise more appropriate
scenarios for the origin of multiple populations in GCs.

Encouraged by this first attempt, here we extend this approach with a systematic
search based on existing abundances of Mg, Ca, and Sc from published analyses of
high-resolution spectra of GC stars, to be compared with a control  sample of
field stars.  In addition to the two notable cases already known (NGC~2419 and
NGC~2808), we  retrieve all the GCs reported in Carretta et al. (2014a). We also
find a few other GCs for which statistically robust evidence of the operation of
proton-capture reactions at very high temperature has been clearly assessed.  A first
scrutiny of their properties in general suggests that these particular classes of
polluters were preferentially at work in massive and/or metal-poor GCs.

The paper is organised as follows: Sect.~2 presents the detection procedure of
excesses in Ca and Sc against a control sample of field stars. Sect.~3  presents
the results of our census for GCs analysed by our group and for the
optical sample from the literature; the new detections are discussed and additional plots are
presented in the appendices. Sect.~4 extends the analysis to the IR, using APOGEE
results, and potassium is added. Sect.~5 discusses a few cases of borderline
detections, and Sect.~6 discusses and summarises our findings.

\section{Setting the stage}

In this section we introduce our procedure to demonstrate possible
excesses of light elements resulting from proton-capture reactions occurring
under extreme conditions in early phases of GC lifetimes. Our approach follows
and expands on the approach used in Carretta and Bragaglia (2019), where we defined
diagnostic plots named detectors of high-temperature H-burning (DOHT). In the
following, we describe our improved procedure in detail, that is, the adopted
reference sample, the different samples of scrutinised GCs, and the statistical
tests we used in the analysis.

\subsection{Reference sample of field stars}
\begin{figure}
\centering
\includegraphics[bb=18 144 300 718, clip, scale=0.8]{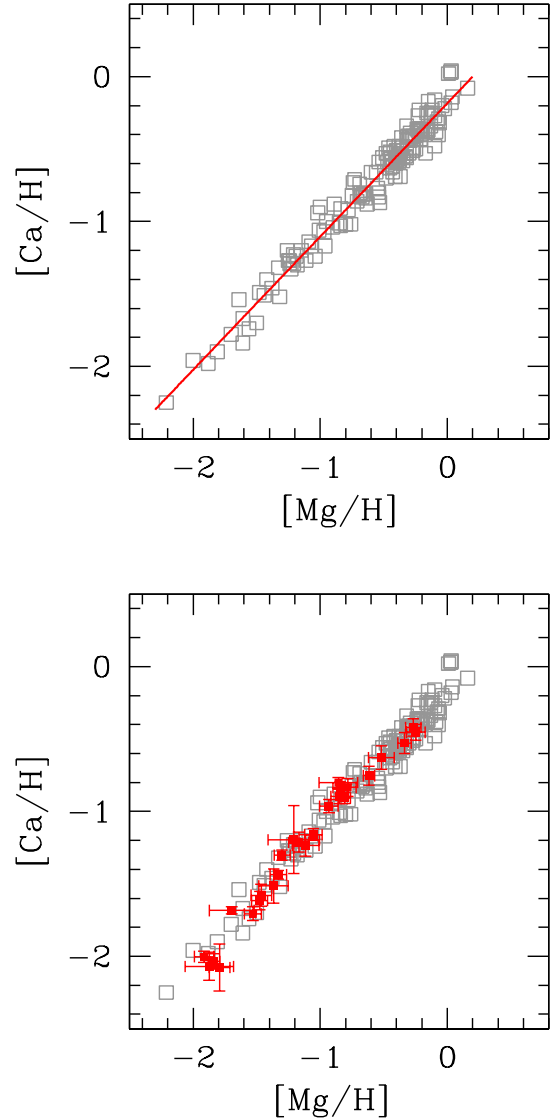}
\caption{Upper panel:[Ca/H] ratios as a function of [Mg/H] for the field
stars in the sample of Gratton et al. (2003). The linear fit to
the data we used to rectify the [Ca/H] values is superimposed. Lower panel: Same as in the
upper panel. The mean values and r.m.s. scatter of GCs in our
golden and silver samples are superimposed (see text).}
\label{f:fitcah}
\end{figure}

As in Carretta and Bragaglia (2019), we tested the presence of signatures of high
temperature burning products in GC multiple populations by comparing their abundance ratio patterns for [Ca/H], [Sc/H], and [Mg/H] with that of field stars because the latter are expected to incorporate only the yields from supernova (SN)
nucleosynthesis (e.g. Gratton et al. 2000, Smith and Martell 2003). Our
preferred reference sample is the one by Gratton et al. (2003).  These stars
cover a range in metallicity from [Fe/H]$=-2.6$  up to [Fe/H]$\sim 0.0$ dex,
encompassing the whole interval of metal abundances spanned by GCs in the Milky
Way (see e.g. Harris 1996, online 2010 edition). Abundances of Mg, Ca, and Sc
were derived for many stars of this sample. The abundance analysis is as
homogeneous as possible with the one for GCs in our golden and silver samples
(from our FLAMES survey, see below) because the adopted line list, reference
solar abundances, and correction for hyperfine structure (HFS) for Sc are
exactly the same.
In Fig.~\ref{f:fitcah} (upper panel) we show the relation between [Ca/H]
and [Mg/H] for stars in Gratton et al. (2003), together with the linear
regression line that we used to subtract the fit from the observed [Ca/H] values
and linearize them for an easier measurement of the Ca excesses.

\subsection{Detection procedure}

In our procedure for detecting excesses of elements produced by H-burning at very
high temperature in GCs, we focused in particular on Ca because this is a
widely studied species, analysed in all the scrutinised GCs. Its abundances are 
often based on several transitions. A possible concern is that the [Ca/Fe] ratio
is overabundant in metal-poor halo stars, either in clusters or in the halo
field (e.g. Gratton et al. 2004), because the contribution of Fe from
type Ia SNe at these low metallicities is reduced, whereas the expected variations in Ca
are likely small. Changes of comparable amounts would be easier to detect if the
involved elements had intrinsically lower abundance, as in the case of K or
Sc. However, this concern is entirely overcome by our differential procedure,
which is tailored to compare field and cluster stars, where the primordial level of Ca is
established by explosive nucleosynthesis in the core-collapse SN phase of
massive stars (e.g. Woosley and Weaver 1995), regardless of the diverse
environment. The initial abundance ratios of $\alpha-$elements 
to iron is determined by the massive star initial mass function (IMF), which is
largely invariant (see e.g. Kordopatis et al. 2015), so that the abundance-scaling relations remain constant for metal-poor stars in the Galaxy. This
evidence is confirmed by the lower panel in Fig.1, where we superimpose  the average values from GCs in our so-called golden
and silver samples on the field star
distribution (see Sections 3.1 and 3.3 below).

On the other hand, Sc is not analysed in all the samples we have access to, and
its derived abundance is affected by additional differences due to the treatment
of corrections for HFS and even of the iron abundance used as reference in the
[Sc/Fe] ratio. It is often unclear whether neutral or singly ionised Fe is
used, and sometimes only the average from both is published, which makes the use of
[Sc/H] ratio more uncertain. However, results based on Sc may be used in support
of the conclusions based on Ca abundances; they are  presented in the appendix.

\begin{figure}
\centering
\includegraphics[scale=0.42]{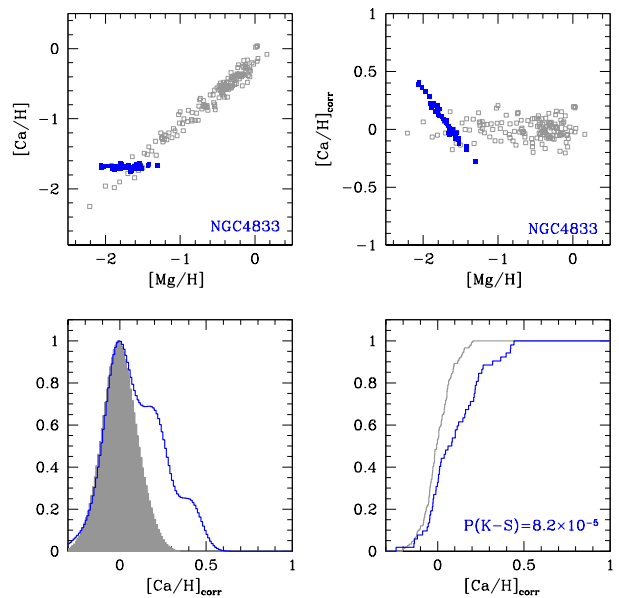}
\caption{Steps of our procedure applied to NGC~4833. Upper left panel: Stars in
NGC~4833 (Carretta et al. 2014a: filled blue squares) superimposed to field
stars (Gratton et al. 2003: empty grey squares) in the [Ca/H-[Mg/H] plane. Upper
right panel: Same stars (and symbols) after subtracting the linear fit in
Fig.~\ref{f:fitcah} from [Ca/H] to obtain linearised values  [Ca/H]$_{\rm
corr}$. Lower left panel: Generalised histograms of the [Ca/H]$_{\rm corr}$
values for field (filled grey area) and GC stars (empty blue area). Lower right
panel: Cumulative distribution of corrected Ca abundances for field and GC
stars. The probability for the two-tail Kolmogorov-Smirnov test is also listed.}
\label{f:f4833}
\end{figure}

To illustrate our procedure, we use NGC~4833. This GC shows the effects we
are looking for more clearly. In Fig.~\ref{f:f4833} the main
steps of the procedure are shown. In the upper left panel the observed
distribution of stars in NGC~4833 from  Carretta et al. (2014a) is superimposed
on the distribution from Gratton et al. (2003) in the [Ca/H] vs [Mg/H] plane. In this
case, the solar reference abundances are the same for the two samples. In the
upper right panel we apply the linear fit in 
Fig.~\ref{f:fitcah} to both samples. In the lower left panel, these ``corrected" [Ca/H]$_{\rm
corr}$ values are plotted as generalised histograms for field and cluster
stars, assuming a typical error of 0.08 dex. We used the complete set of
field stars after verifying that the ratio [Ca/H]$_{\rm corr}$ runs flat,
without modification of average and dispersion. In this step, the two samples
were aligned using the mode of the distributions; in practice, the GC peak was
shifted to correspond to the maximum of the field star distribution. This also
accounts for residual small differences that are due to the adoption of a different
temperature scale in the analyses, for instance. 

Simple visual inspection is clearly enough in this example (as well as in most
of the cases we analysed) to detect an evident excess of Ca in a fraction of GC
stars, excess that is absent in the field stars. To better quantify these effects,
we applied a two-sample Kolmogorov-Smirnov (K-S) test to each GC under the null
hypothesis that the two sample populations (GC and field stars) are drawn from
the same parent distribution. In the case of NGC~4833, the cumulative
distributions of the corrected [Ca/H] values are compared in the lower right
panel of Fig.~\ref{f:f4833}, where the probability for the two-tail K-S test is
also indicated. For NGC~4833, the result of the test unambiguously shows that
the two distributions are not extracted from the same parent population. A
fraction of stars of this cluster shows an excess of Ca above the level
established by SNe in field halo stars with similar metallicity. The same
procedure can also be used also with Sc, if available; for NGC~4833 the Sc
distribution confirms the results obtained with Ca (the K-S test indicates a high
significance,  with P(K-S)=$4.3\times10^{-5}$.

We also computed the number of outliers in the GC linearised distributions, 
defined as those GC stars whose corrected [Ca/H]$_{\rm corr}$ ratios exceed the
3$\sigma$ range with respect to the average for field stars (where the last
value is zero by definition in our procedure). We adopted $r.m.s=0.084$ dex for
[Ca/H] from Gratton et al. (2003). This value is only slightly above the
error in Ca because of the uncertainties in the analysis. We verified that this is also
a typical and reasonable value to be used for most of the literature Ca
abundances. In the example above, NGC~4833 has 8 stars (of the 52 in our
sample, i.e. 15\%) that exceed the 3$\sigma$ limit.

The same steps can be applied to all other GCs (see next section). Whenever
possible (and explicitly published), we used the solar reference  abundances from
the original studies to remove a source of offset by  shifting the abundances
on the solar reference scale used by Gratton  et al. (2003) and in our FLAMES
survey of GCs (i.e. $\log$ $\epsilon$(Fe)=7.54, $\log$ $\epsilon$(Mg)=7.43, 
$\log$ $\epsilon$(Ca)=6.27, and $\log$ $\epsilon$(Sc)=3.13).  The
distribution peak alignment (Fig.~\ref{f:f4833}) takes care of residual small
offsets caused by differences in the adopted solar abundances.

\begin{figure*}
\centering
\includegraphics[scale=0.70]{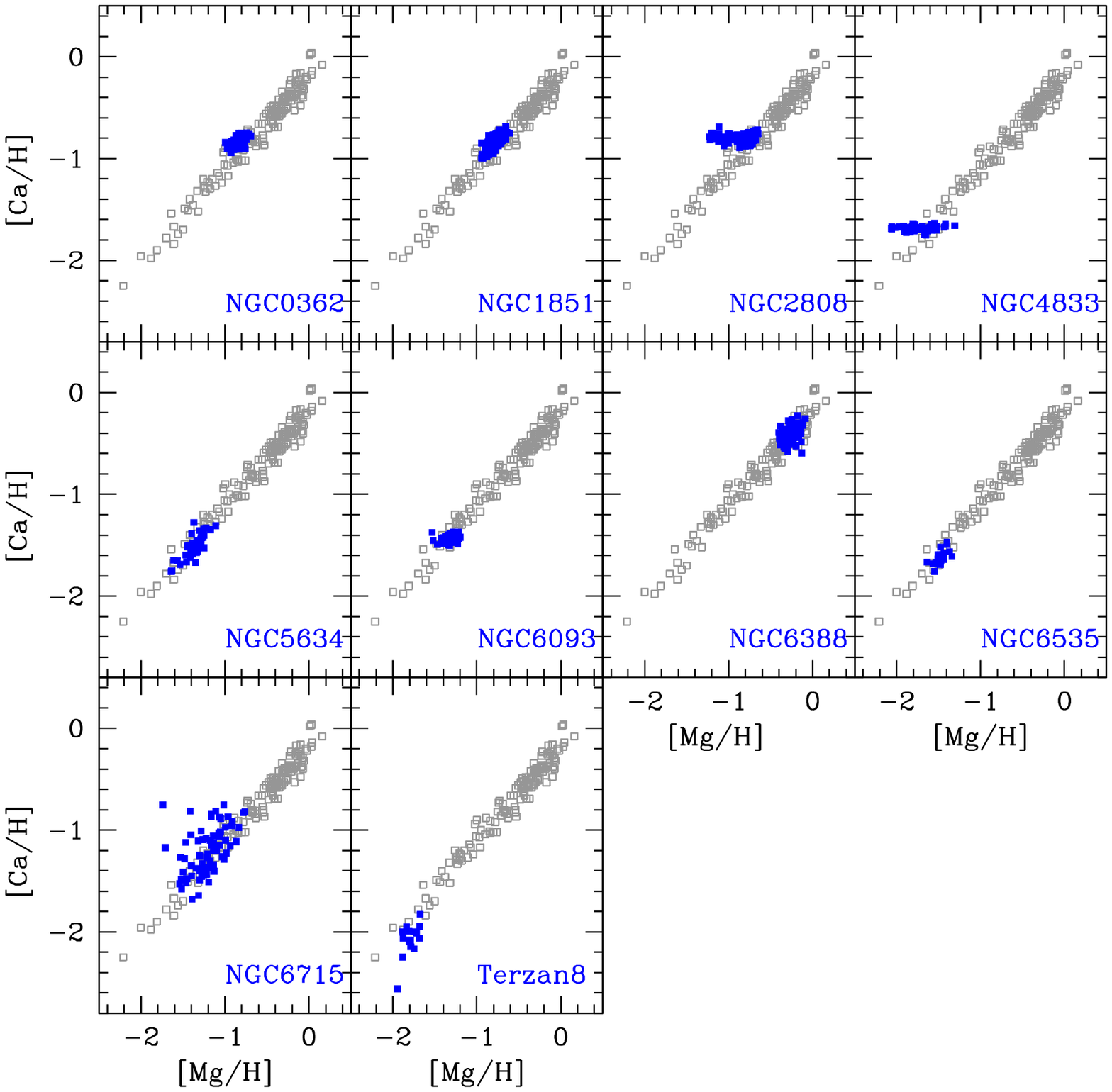}
\caption{Observed [Ca/H] ratios as a function of [Mg/H] ratios for the GCs in
our golden sample (filled blue squares) compared to the field stars in
Gratton et al.  (2003: empty grey squares).}
\label{f:goldenA}
\end{figure*}

\setcounter{table}{0}
\begin{table*}
\centering
\setlength{\tabcolsep}{1mm}
\caption{Adopted abundance analyses }
\begin{tabular}{llrl|llrl}
\hline
\hline
GC      & Code        &Nstar  & Reference &GC      & Code        &Nstar  & Reference \\
        & (study)     &       &           &        & (study)     &       &             \\
\hline
\multicolumn{8}{c}{golden sample}      \\       
  0362  &  C13       &  86   & Carretta et al. (2013b)        & 6093  &  C15A        &  80   & Carretta et al. (2015)       \\  
  1851  &  C11       & 123   & Carretta et al. (2011)         & 6388  &  CB20        & 184   & Carretta \& Bragaglia (2020) \\  
  2808  &  C15       & 139   & Carretta (2015)                & 6535  &  Br17        &  24   & Bragaglia et al. (2017)      \\  
  4833  &  C14A      &  52   & Carretta et al. (2014a)        & 6715  &  C10A        &  74   & Carretta et al. (2010a)      \\  
  5634  &  C17       &  42   & Carretta et al. (2017)         & Ter8  &  C14B        &  16   & Carretta et al. (2014b)      \\  
\multicolumn{8}{c}{silver sample}      \\       
  0104  &  C09       &  11   & Carretta et al. (2009a,2010b)   & 6218  &  C09        &  11   & Carretta et al. (2009a,2010b) \\ 
  0288  &  C09       &  10   & Carretta et al. (2009a,2010b)   & 6254  &  C09        &  14   & Carretta et al. (2009a,2010b) \\ 
  1904  &  C09       &  10   & Carretta et al. (2009a,2010b)   & 6397  &  C09        &  13   & Carretta et al. (2009a,2010b) \\ 
  3201  &  C09       &  13   & Carretta et al. (2009a,2010b)   & 6752  &  C09        &  14   & Carretta et al. (2009a,2010b) \\ 
  4590  &  C09       &  13   & Carretta et al. (2009a,2010b)   & 6809  &  C09        &  14   & Carretta et al. (2009a,2010b) \\ 
  5904  &  C09       &  14   & Carretta et al. (2009a,2010b)   & 6838  &  C09        &  11   & Carretta et al. (2009a,2010b) \\ 
  6121  &  C09       &  14   & Carretta et al. (2009a,2010b)   & 7078  &  C09        &  13   & Carretta et al. (2009a,2010b) \\ 
  6171  &  C09       &   5   & Carretta et al. (2009a,2010b)   & 7099  &  C09        &  10   & Carretta et al. (2009a,2010b) \\ 
\multicolumn{8}{c}{literature sample}      \\   
  0104  &  C04       &  10   & Carretta et al. (2004)         & 6366  &  Puls18     &    8   & Puls et al. (2018)            \\ 
  0104  &  THYG14    &  13   & Thygesen et al. (2014)         & 6397  &  Lind11     &  21   & Lind et al. (2011)     \\ 
  0288  &  SHE00     &  13   & Shetrone \& Keane (2000)       & 6402  &  J19         &  32   & Johnson et al. (2019)         \\ 
  0362  &  SHE00     &  12   & Shetrone \& Keane (2000)       & 6426  &  H17         &   4   & Hanke et al. (2017)           \\ 
  2419  &  CK12      &  13   & Cohen \& Kirby (2012)          & 6440  &  Mun17      &    7   & Munoz et al. (2017)           \\ 
  2419  &  Mu12      &  47   & Mucciarelli et al. (2012)      & 6441  &  GR06-07    &  28   & Gratton et al. (2006,2007)    \\ 
  2808  &  M17       &   7   & Marino et al. (2017)           & 6522  &  B09         &   8   & Barbuy et al. (2009)          \\ 
  3201  &  Mun13     &   8   & Munoz et al. (2013)            & 6522  &  N14         &   8   & Ness et al. (2014)            \\ 
  3201  &  GW98      &  17   & Gonzalez \& Wallerstein (1998) & 6528  &  Mun18      &    7   & Munoz et al. (2018)           \\ 
  3201  &  Mag18     &   7   & Magurno et al. (2018)          & 6553  &  J14         &  12   & Johnson et al. (2014)         \\ 
  3201  &  M19       &  18   & Marino et al. (2019)           & 6553  &  Tang17     &  10   & Tang et al. (2017)     \\ 
  4147  &  V16       &   5   & Villanova et al. (2016)        & 6558  &  B18         &   3   & Barbuy et al. (2018)          \\ 
  4372  &  SR15      &   7   & San Roman et al. (2015)        & 6558  &  B07         &   5   & Barbuy et al. (2007)          \\ 
  4590  &  S15       &  25   & Schaeuble et al. (2015)        & 6569  &  J18         &  19   & Johnson et al. (2018)         \\ 
  4590  &  Lee05     &   7   & Lee et al. (2005)              & 6569  &  VOR11      &  11   & Valenti et al. (2011)          \\ 
  4833  &  ROE15     &  15   & Roederer \& Thompson (2015)    & 6626  &  V17         &  17   & Villanova et al. (2017)       \\ 
  5139  &  NDC95     &  40   & Norris \& Da Costa (1995)      & 6656  &  M11         &  35   & Marino et al. (2011)          \\ 
  5139  &  V10       &  28   & Villanova et al. (2010)        & 6681  &  OM17        &   9   & O'Malley et al. (2017)        \\ 
  5272  &  CM05A     &  13   & Cohen \ Melendez (2005a)       & 6723  &  CR19        &  11   & Crestani et al. (2019)        \\ 
  5272  &  SNE04     &  23   & Sneden et al. (2004)           & 6752  &  Y05         &  38   & Yong et al. (2005)            \\ 
  5286  &  M15       &   7   & Marino et al. (2015)           & 6752  &  GR05        &   7   & Gratton et al. (2005)         \\ 
  5466  &  L15       &   3   & Lamb et al. (2015)             & 6809  &  Rain19     &  11   & Rain et al. (2019)     \\ 
  5694  &  Mu13      &   6   & Mucciarelli et al. (2013)      & 6838  &  RC02        &  24   & Ramirez \& Cohen (2002)       \\ 
  5824  &  ROE16     &  26   & Roederer et al. (2016)         & 6864  &  K13         &  16   & Kacharov et al. (2013)        \\ 
  5897  &  KMW14     &   7   & Koch \& McWilliam (2014)       & 6934  &  M18         &   4   & Marino et al. (2018)          \\ 
  5904  &  RC03      &  17   & Ramirez \& Cohen (2003)        & 7006  &  Kra98      &    6   & Kraft et al. (1998)           \\ 
  5904  &  Lai11     &  17   & Lai et al. (2011)              & 7078  &  Sne97      &  18   & Sneden et al. (1997)           \\ 
  5927  &  M-G18     &   7   & Mura-Guzman et al. (2018)      & 7089  &  Y14B        &  13   & Yong et al. (2014b)           \\ 
  5986  &  J17A      &  24   & Johnson et al. (2017a)         & 7492  &  CM05B      &    4   & Cohen \& Melendez (2005b)     \\ 
  6121  &  M08       & 105   & Marino et al. (2008)           & Arp2  &  MWM08      &    2   & Mottini et al. (2008)         \\ 
  6121  &  I99       &  24   & Ivans et al. (1999)            & E3    &  Mon18      &    2   & Monaco et al. (2018)          \\ 
  6121  &  M17       &  17   & Marino et al. (2017)           & F1758 &  V19         &   9   & Villanova et al. (2019)       \\ 
  6121  &  V11       &  19   & Villanova \& Geisler (2011)    & Gaia1 &  K18         &   4   & Koch et al. (2018)            \\  
  6205  &  CM05A     &  21   & Cohen \& Melendez (2005a)      & HP1   &  B16         &   8   & Barbuy et al. (2016)          \\  
  6205  &  Kra97     &  18   & Kraft et al. (1997)            & Pal1  &  S11         &   5   & Sakari et al. (2011)$^a$      \\  
  6218  &  J06       &  11   & Johnson \& Pilachowski (2006)  & Pal3  &  K09         &   4   & Koch et al. (2009)            \\  
  6229  &  J17B      &   8   & Johnson et al. (2017b)         & Pal12 &  Coh04      &    4   & Cohen (2004)                  \\  
  6266  &  Y14A      &   7   & Yong et al. (2014a)            & Pal14 &  CCG12      &    9   & Caliskan et al. (2012)        \\  
  6273  &  J17C      &  28   & Johnson et al. (2017c)         & Pal15 &  K19         &   3   & Koch et al. (2019b)           \\  
  6352  &  F09       &   9   & Feltzing et al. (2009)         & Rup106&  V13         &   9   & Villanova et al. (2013)       \\  
  6362  &  Mas17     &  11   & Massari et al. (2017)          & Ter7  &  S07         &   5   & Sbordone et al. (2007)        \\  
  6366  &  J16       &  13   & Johnson et al. (2016)          & Ter8  &  MWM08      &    3   & Mottini et al. (2008)         \\  
\hline
\end{tabular}
\begin{list}{}{}
\item[$^a$] One star is repeated, from Monaco et al. (2011)
\end{list}
\label{t:tab1}
\end{table*}

\section{Analysis and results}

The procedure described in the previous section for NGC~4833 was applied to all
GCs for which we found published abundances of Fe, Mg, and Ca (and Sc whenever
possible: see Appendix B) based on high-resolution spectra. The GCs and the
(sometimes multiple) abundance analyses used in the present study are listed in
Tab~\ref{t:tab1}.  For each GC we list the NGC number or name, a unique
alphanumeric code identifying  each individual study in each different GC in the
plots shown both in the main paper body and in the appendices, the number of
stars analysed in each GC, and the reference of the paper.  

The GCs in Table~\ref{t:tab1} are grouped into three separated sub-samples.  Our
FLAMES survey conducted to study the O-Na anti-correlation in GCs, described in Carretta
et al. (2006, 2010c), allowed us to analyse an unprecedented sample of more than
2800 giants in 27 GCs in an extremely homogeneous way, with high-resolution
FLAMES spectra.  Not all stars have the full set of abundances derived for all
elements, but  in our so-called $\text{golden}$ sample, we collected 9 GCs with
published abundances of Fe, Mg, Ca, and Sc that have been derived from either UVES or
GIRAFFE  spectra. To these we also added NGC~6388, whose data are still
unpublished (paper in preparation). A total of 817 stars with homogeneously
derived abundances in 10 GCs constitutes our prime sample.

Second, our $\text{silver}$ sample consists of 16 GCs in which Fe, Mg, and Ca
abundances were derived from UVES spectra in Carretta et al. (2009a) and
Carretta et al. (2010b) for a total of 190 giants. In this case, while the high
degree of  homogeneity is preserved, the sample is limited to a maximum of 14
stars in each GC because only few UVES fibres are available. This subset represents
our second-best sample for its homogeneity, although the rather small samples
are not ideal.

Finally, in the third group we include all the samples retrieved in the literature
with abundances of Fe, Mg, and Ca based on high-resolution optical spectroscopy
(plus one near-IR for NGC~6553). The number of stars in each study varies strongly,
from 2 (Arp~2 and E3) to more than 100 (NGC~6121). The near-IR results in the
latest APOGEE data release are presented and discussed in Sect.~4.

The results of the K-S test on Ca abundance distribution for all studies
(labelled by their code) of all GCs (indicated by NGC number or name) are listed
in Tab.~\ref{t:tab2}, together with the number of outliers in  the GC
[Ca/H]$_{\rm corr}$ ratios. The literature sample in Tab.~\ref{t:tab2} is split
into two subsets, according to the number of stars examined in each GC (more and
fewer than 5). In Appendix A we show the observed [Ca/H] ratios as a function of
[Mg/H] for all literature samples.  

\setcounter{table}{1}
\begin{table*}
\centering
\caption{Results for [Ca/H]-[Mg/H]}
\begin{tabular}{lllr|lllr}
\hline
\hline
GC   & Code & prob  &  outliers &GC   & Code & prob  &  outliers   \\
     &      & K-S   &           &     &      & K-S   &            \\
\hline
\multicolumn{8}{c}{golden sample}      \\       
  0362   & C13       & 0.172                  &    0 &   6093   & C15A      & 0.115                   &    0  \\       
  1851   & C11       & 0.021                  &    0 &   6388   & CB20      & 0.232                   &    0  \\       
  2808   & C15       & 2.7$\times 10^{-3}$    &   22 &   6535   & Br17      & 0.245                   &    0  \\   
  4833   & C14A      & 8.2$\times 10^{-5}$    &    8 &   6715   & C10A      & 0.022                   &   12  \\ 
  5634   & C17       & 0.549                  &    0 &   Ter8   & C14B      & 0.528                   &    0  \\       
\multicolumn{8}{c}{silver sample}      \\       
  0104   & C09       & 0.369                  &    0 &   6218   & C09       & 0.340                   &    0  \\ 
  0288   & C09       & 0.166                  &    0 &   6254   & C09       & 0.506                   &    0  \\ 
  1904   & C09       & 0.284                  &    0 &   6397   & C09       & 0.247                   &    0  \\ 
  3201   & C09       & 0.284                  &    0 &   6752   & C09       & 0.500                   &    0  \\ 
  4590   & C09       & 0.596                  &    0 &   6809   & C09       & 0.622                   &    0  \\ 
  5904   & C09       & 0.694                  &    0 &   6838   & C09       & 0.828                   &    0  \\ 
  6121   & C09       & 0.435                  &    0 &   7078   & C09       & 0.575                   &    2  \\ 
  6171   & C09       & 0.258                  &    0 &   7099   & C09       & 0.284                   &    0  \\ 
\multicolumn{8}{c}{literature (N$_{\rm star} \ge 5$) }  \\  
  0104   & C04       & 0.124                  &   0 &   6266   & Y14A      & 0.258                   &   0  \\ 
  0104   & THYG14    & 0.762                  &   0 &   6273   & J17C      & 0.407                   &   1  \\ 
  0288   & SHE00     & 0.695                  &   0 &   6352   & F09       & 0.219                   &   0  \\ 
  0362   & SHE00     & 0.177                  &   0 &   6362   & Mas17     & 0.442                   &   0  \\ 
  2419   & CK12      & 1.2$\times 10^{-3}$    &   5 &   6366   & J16       & 0.524                   &   0  \\ 
  2419   & Mu12      & 5.7$\times 10^{-6}$    &  17 &   6366   & Puls18    & 0.663                   &   2  \\ 
  2808   & M17       & 0.956                  &   0 &   6397   & Lind11    & 0.373                   &   0  \\ 
  3201   & Mun13     & 0.852                  &   0 &   6402   & J19       & 4.7$\times 10^{-3}$    &   3  \\ 
  3201   & GW98      & 0.890                  &   1 &   6440   & Mun17     & 0.014                   &   3  \\ 
  3201   & Mag18     & 0.277                  &   1 &   6441   & GR06-07   & 3.811$\times 10^{-5}$  &   7  \\ 
  3201   & M19       & 0.086                  &   0 &   6522   & B09       & 0.121                   &   0  \\ 
  4147   & V16       & 0.552                  &   0 &   6522   & N14       & 0.978                   &   0  \\ 
  4372   & SR15      & 0.723                  &   0 &   6528   & Mun18     & 0.863                   &   0  \\ 
  4590   & S15       & 0.023                  &   1 &   6553   & J14       & 0.047                   &   1  \\ 
  4590   & Lee05     & 0.757                  &   1 &   6553   & Tang17    & 0.302                   &   0  \\ 
  4833   & ROE15     & 3.2$\times 10^{-3}$    &   7 &   6558$^a$  & B07+B18   & 0.191                   &   0  \\ 
  5139   & NDC95     & 9.8$\times 10^{-5}$    &  10 &   6569   & J18       & 0.089                   &   0  \\ 
  5139   & V10       & 1.3$\times 10^{-4}$    &   9 &   6569   & VOR11     & 0.163                   &   0  \\ 
  5272   & CM05A     & 0.610                  &   0 &   6626   & V17       & 0.787                   &   0  \\ 
  5272   & SNE04     & 0.428                  &   1 &   6656   & M11       & 0.278                   &   1  \\ 
  5286   & M15       & 0.102                  &   0 &   6681   & OM17      & 0.893                   &   0  \\ 
  5694   & Mu13      & 0.330                  &   0 &   6723   & CR19      & 0.163                   &   0  \\ 
  5824   & ROE16     & 3.2$\times 10^{-4}$    &   7 &   6752   & Y05       & 0.436                   &   0  \\ 
  5897   & KMW14     & 0.899                  &   0 &   6752   & GR05      & 0.867                   &   0  \\ 
  5904   & RC03      & 0.347                  &   0 &   6809   & Rain19    & 0.642                   &   0  \\ 
  5904   & Lai11     & 0.989                  &   0 &   6838   & RC02      & 0.086                   &   0  \\ 
  5927   & M-G18     & 0.363                  &   0 &   6864   & K13       & 0.713                   &   0  \\ 
  5986   & J17       & 8.9$\times 10^{-3}$    &   4 &   7006   & Kra98     & 0.310                   &   0  \\ 
  6121   & M08       & 0.575                  &   0 &   7078   & Sne97     & 0.109                   &   4  \\ 
  6121   & I99       & 0.542                  &   0 &   7089   & Y14B      & 0.252                   &   0  \\ 
  6121   & M17       & 0.229                  &   1 &   F1758  & V19       & 0.313                   &   0  \\ 
  6121   & V11       & 0.375                  &   0 &   HP1    & B16       & 0.776                   &   0  \\ 
  6205   & CM05A     & 0.036                  &   4 &   Pal1   & S11       & 0.342                   &   0  \\ 
  6205   & Kra97     & 0.090                  &   0 &   Pal14  & CCG12     & 0.015                   &   0  \\ 
  6218   & J06       & 0.005                  &   0 &   Rup106 & V13       & 0.695                   &   0  \\ 
  6229   & J17B      & 0.806                  &   0 &   Ter7   & S07       & 0.547                   &   0  \\ 
\multicolumn{8}{c}{literature (N$_{\rm star} < 5$) }      \\    
  5466   & L15       & 0.688                  &   0 &   Gaia1  & K18       & 0.826                   &   0  \\
  6426   & H17       & 0.925                  &   0 &   Pal3   & K09       & 0.097                   &   0  \\ 
  6934   & M18       & 0.543                  &   0 &   Pal12  & Coh04     & 0.861                   &   0  \\ 
  7492   & CM05B     & 0.603                  &   0 &   Pal15  & K19       & 0.772                   &   0  \\ 
  Arp2   & MWM08     & 0.542                  &   0 &   Ter8   & MWM08     & 0.752                   &   0  \\ 
  E3     & Mon18     & 0.465                  &   1 &          &           &                        &      \\     

\hline
\end{tabular}
\begin{list}{}{}
\item[$^a$] The final sample for this GC was obtained by combining two analyses
from the same group.
\end{list}
\label{t:tab2}
\end{table*}

\subsection{Golden sample}

The observed distribution of stars in GCs of our golden sample is shown in the
plane [Ca/H]-[Mg/H] in Fig.~\ref{f:goldenA}. Visual inspection shows that the
cluster stars are clearly superimposed on the locus
populated by field stars in most cases, with three exceptions: NGC~2808,
NGC~4833, and NGC~6715 (M~54).

NGC~2808 is a case study for correlations and anti-correlations involving
products forged in H-burning in conditions of very high temperature. Carretta
(2015) showed that among the multiple populations in this GC, the Sc abundances
are anti-correlated with those of Mg and are well correlated with Si and Ca
abundances. At high statistic significance, all these relations indicate that
a fraction of stars in NGC~2808 is contaminated by nuclear matter processed at
very high temperature in proton-capture reactions.  Figure~\ref{f:goldenA} is
another way to present these observations. It also immediately shows the
excess of this production with respect to field stars where Si and Ca originated
only in SN nucleosynthesis.

To this template case we now add NGC~4833, which seems to be an analogy on a
smaller scale. This is expected because the original analysis by Carretta
et al. (2014a) has clearly shown that this GC stands out in the [Ca/Mg] versus
[Ca/H] plane, another diagnostic plot introduced in  Carretta et al. (2013a) to
single out GCs with an excess of Ca and/or a high Mg depletion. A group of Mg-poor stars was also found in a more limited sample by Roederer
and Thompson (2015).

The third detection is represented by NGC~6715 (M~54). Some giants  clearly
show excesses of Ca with respect to field stars over the range in metallicity
covered by this former nuclear cluster of the dwarf galaxy Sagittarius.

The probability (p-value) associated with the two-tail K-S test is listed in the
first part of Tab.~\ref{t:tab2} for the GCs of the golden sample. We adopted the
usual convention that a p-value $<0.05$ means that there is sufficient evidence
to reject the null hypothesis that the GC and field star samples are extracted
from the same parent population. The low values we found for NGC~2808, NGC~4833,
and NGC~6715 confirm the detected difference
between field and cluster stars on statistically robust grounds. In all these three samples, a fraction of
15-16\% of the stars is classified as outliers, following the criteria described
above. The case for NGC~1851, with a p-value below 0.05 and no outlier above the
3$\sigma$ threshold (as is immediately clear from the visual inspection) is
discussed in the next subsection.

These three detections are also mirrored, albeit not so clearly, in the
[Sc/H] versus [Mg/H] plane (see Appendix B, Figure B.1). The K-S test and the numerous outliers in the Sc distribution also strengthen our 
findings based on Ca.

\subsection{Proof-of-concept case: NGC~1851}
The particular case of NGC~1851 appears to show a discrepancy: while
in Fig.~\ref{f:goldenA} the cluster stars are well superimposed on the  field
stars of similar metallicity, the K-S test in Tab.~\ref{t:tab2} formally
indicates that the probability that we can reject the null hypothesis is low
(i.e. that the samples are extracted from the same parent population). 

\begin{figure}
\centering
\includegraphics[scale=0.42]{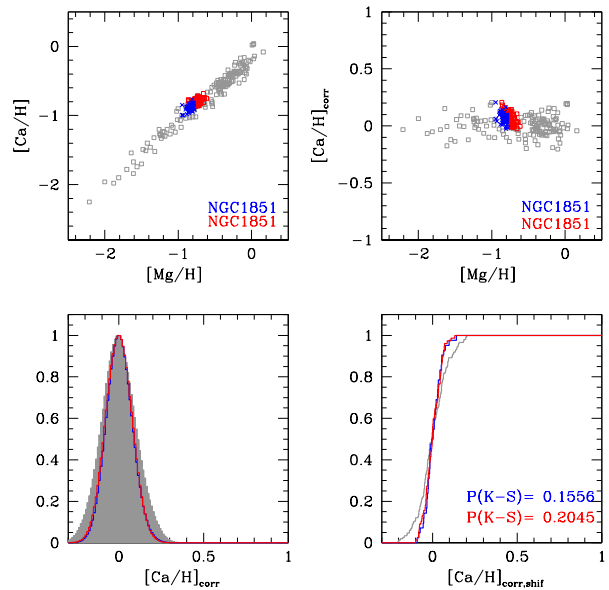}
\caption{Analysis for the detection and test of the Ca excess in NGC~1851: the whole
procedure is now applied separately to the two stellar components (red and blue
symbols and lines) found by Carretta et al. (2011) in this cluster.}
\label{f:concept18}
\end{figure}

The explanation is that two different
sub-populations coexist in this GC. Carretta et al. (2011) showed that the difference in Fe alone 
is too small for a clear distinction of these two populations;  however, the
separation is evident in a classification scheme with a combination of Fe and
Ba, where the latter is much more abundant in the more metal-rich component. In
other words, NGC~1851 appears to be a smaller-scale example of the so-called iron
complex GCs, such as NGC~6656 (M~22: e.g. Marino et al. 2011), NGC~6273 (M~19:
Johnson et al. 2015, 2017c), and NGC~5824 (Roederer et al. 2016).  In these
GCs, a spread in Fe is accompanied by a dispersion in $s-$process elements,
whose content is higher in the most metal-rich fraction. Carretta et al.
(2011)found that each component in NGC~1851 hosts a fully developed O-Na
anti-correlation by its own, but the two populations also have different
levels of [Ca/H], [Mg/H], [Si/H] with a statistically high level  of confidence.

Because we used both Ca and Mg in our tests, we repeated our procedure for
NGC~1851 and separated the two components as found and defined in Carretta et al.
(2011). The results are shown in Fig.~\ref{f:concept18}. Each individual
component does not seem to be significantly different from the field stars, as
established by the probabilities of the K-S test. On the other hand, the two components are clearly segregated and clustered around two
slightly different values in the entire sample. This in turn results in a less than perfect match
with the underlying field stars, and this spurious signal is most probably what 
is intercepted by the K-S test based on the total sample of NGC~1851.
The discussion in this section can be also used as a sanity check of our 
procedure and as a proof of concept that it efficiently works in detecting the
excesses we are interested in. 

\subsection{Silver sample}

In terms of homogeneity, our silver sample GCs are next because their stars
were analysed with the same tools (line lists, solar reference abundances, HFS
corrections, and abundance analysis code) as the stars in the golden sample and  in
the field. The only relevant difference is that these stars were observed with
the dedicated UVES-FLAMES fibres, hence the number of stars in each cluster is
smaller\footnote{We recall that for each FLAMES plate, there are about 130
fibres for the GIRAFFE and only 8 for the UVES spectrographs.}. As a compensation, the number of transitions for Fe, Ca, and Sc
is higher because the spectral coverage of UVES spectra is larger, providing more
robust estimates of the abundances.

\begin{figure*}
\centering
\includegraphics[scale=0.70]{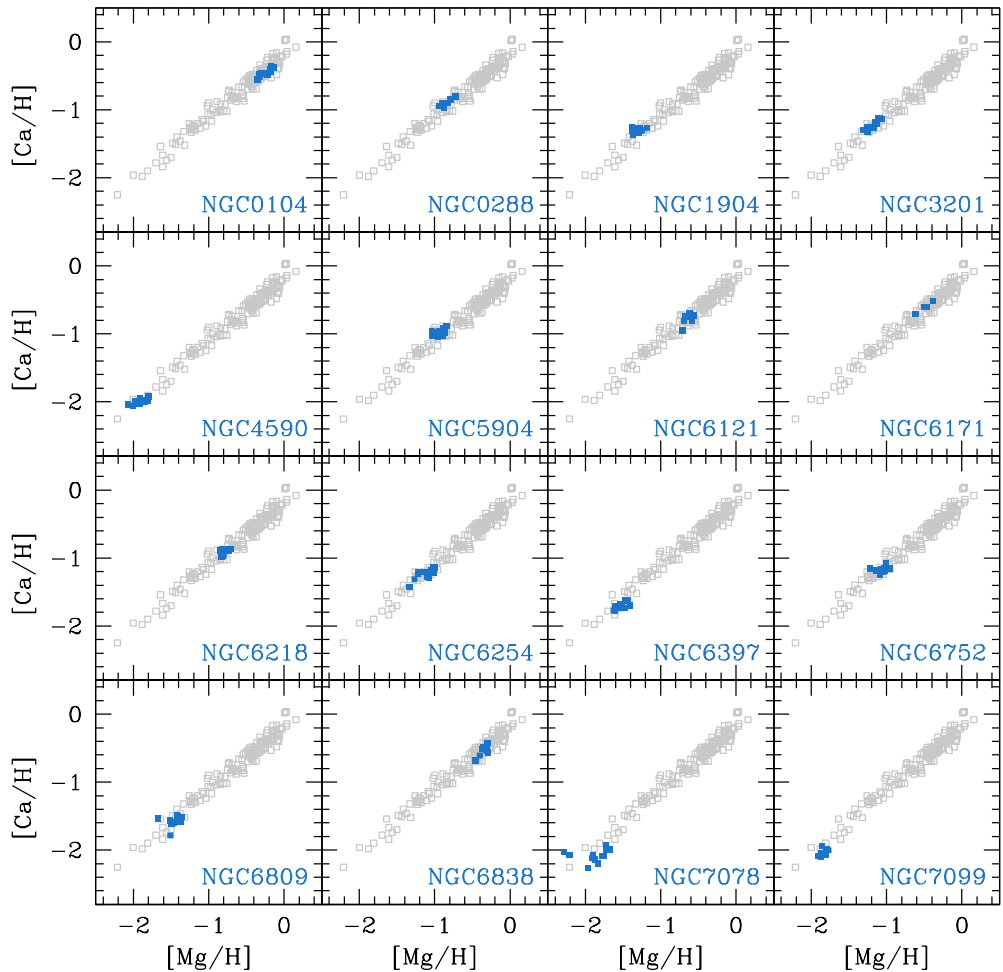}
\caption{Observed [Ca/H] ratios as a function of [Mg/H] ratios for the  GCs in
our silver sample (filled blue squares) compared to the field stars  in
Gratton et al. (2003), shown in grey.}
\label{f:silverA}
\end{figure*}

The diagnostic plots showing the observed distributions of GC stars and field
stars  in the [Mg/H], [Ca/H] plane are presented in Fig.~\ref{f:silverA}. The
p-values for the K-S test are listed in the second part of  Table~\ref{t:tab2}: 
no GC in this sample shows evidence of excess of Ca with respect to field stars.

Interestingly, the GCs in this set are the same as were used in Carretta et al. (2009a)
to explore the various relations among the light elements O, Na, Mg, Al, and Si whose
abundances are potentially changed in proton-capture reactions involving a large temperature interval for nuclear burning. The only significant slopes in the
Si-Al correlation were found for NGC~2808, NGC~6388, NGC~6752, and NGC~7078
(Carretta et al. 2009a, figure 10). Translated into temperature, this finding
means that FG polluters were only in these GCs able to reach the 65 MK 
threshold (Arnould et al. 1999), providing the right environment for leakage on
$^{28}$Si in the Mg-Al cycle (see Karakas and Lattanzio 2003).

Carretta and Bragaglia (2019) have examined the cases of NGC~2808 and
NGC~6388, which stimulated the census of the present work. By applying the same
line of reasoning, we may conclude that in NGC~6752 we can also pinpoint a rather
narrow range for the operating temperatures of the polluters, as  we did in
NGC~6388: high enough to modify  Si, but not high enough to touch on heavier
elements. Mucciarelli et al. (2017) offered a nice support to this
conclusion because they found that in NGC~6752 the results are formally
compatible with a null spread in K abundance.

The verdict on NGC~7078 (M~15) is more uncertain. Two outliers, at slightly more
than 3$\sigma$ in [Ca/H]$_{\rm corr}$, might indicate a low excess in Ca, in
agreement also with the diagnostic plot [Ca/Mg] versus [Ca/H] shown in Carretta et
al. (2014a). However, the K-S test does not formally support this conclusion,
and  we cannot currently strongly state that stars in NGC~7078 are genuinely
different from the distribution of field stars. Unfortunately, the various
literature studies  do not help solve this issue either (see below). It would be
interesting to make a survey of K abundances in this GC:   the  K~{\sc i}
resonance line at 7698.98~\AA\ should be still strong enough to be reliably
measured in giant stars of such a low-metallicity cluster as well.

\subsection{Literature sample}

To enlarge the set of GCs, we  collected literature samples in which Mg, Ca, and
possibly Sc were derived from high-resolution spectra (in this section we consider optical spectra, APOGEE is discussed in Sect.~4). Multiple studies of the
same cluster were considered to improve our chance of detecting stars with a
clear excess in  Ca and/or Sc.  Only in two cases did we try to combine different
analyses for the same cluster to reach a more statistically meaningful number
of stars.  However, for NGC~5466 the optical sample by L15 and the near-IR from
APOGEE (Meszaros et al. 2020, see next section) display an offset, therefore we did
not combine them.  Conversely, for NGC~6558, we summed the samples B07
and B18 because they are from the same group; we only took unique
stars. 

Whenever possible, abundances were corrected for offsets due to solar reference
abundances with respect to those in Gratton et al. (2003). However, the
procedure that normalises the GC stars to the  bulk of field star in
[Ca/H]$_{\rm corr}$ helps to smooth out residual small differences due to the
analysis.

In Appendix A (figures from A1 to A10) we show the observed distributions of
[Ca/H] and [Mg/H] superimposed on the field stars by Gratton et al. (2003).  In
the panels, each GC study is identified with the alphanumeric code in
Tab.~\ref{t:tab1} and  Tab.~\ref{t:tab2}.  NGC~2419 is  plotted with a
different scale that is required to show the full extent of its huge spreads in Ca and
Mg (and Sc, see Appendix B). This cluster clearly stands out with its large
alterations in these elements, as was discovered in Cohen and Kirby (2012)
and Mucciarelli et al. (2012).

The results of the two-tail K-S test are listed in Tab.~\ref{t:tab2}.  From
these tests, a (moderate) number of GCs show a clear signature of the
operation of proton-capture reactions at very high temperature, meaning that the
low p-values allow us to safely state that the difference between the two
distributions (field versus GC stars) is real. In addition, if different studies
exist for the same GC, all the corresponding statistical tests  give results that agree with each other.

In addition to NGC~2419, we detected such a clear signal in NGC~4833, NGC~5139 ($\omega$
Cen), NGC~5824, NGC~5986, and NGC~6402, all GCs with a large number of stars 
analysed. The analysis of NGC~4833 by Roederer et al. (2015) independently
confirms the result we obtained for this GC in our golden sample. The excess in
$\omega$ Centauri, the most massive GC in the Milky Way, is detected by two
independent studies (which become three when the APOGEE
contribution is included, see next section). In its case, the large variations are 
clearly evident from the Mg and Ca distribution (see Fig.~A.2).

Moreover, we found a few GCs where the conclusion is unclear: either
because different analyses for the same cluster provide different answers, or
because the sample in the GC shows a large dispersion that is not skewed only 
towards excess values. In the latter case, we cannot be sure that the effect is
astrophysical and not simply due to large errors associated with the abundance
analysis. 

The list of these GCs includes NGC~4590 (M~68),  NGC~6205 (M~13), NGC~6218
(M~12), NGC~6440, NGC~6441, NGC~6553, and Pal~14.  We discuss these objects
below, but in the following we do not consider the presence of matter
produced at very high temperature H-burning in these GCs conclusive. Two of them are also
in our silver sample (NGC~6218 and  NGC~6838), where they do not show any Ca
excess.

Globular clusters for which only a limited number of stars ($<5$) is available are
plotted in different colours  in the figures in Appendices A and B and are listed in
the last lines of Tab.~\ref{t:tab2}. The K-S test shows that the cluster and
field star distributions do not differ for any of these GCs. The soundness of
this conclusion is limited by the small available samples.   However, for
Ter~8  the result is rather well assessed on the basis of the  larger sample
analysed by Carretta et al. (2014b). Because of the small
numbers involved, the lack of evidence  for the remaining GCs must be considered  only as an indication, to
be confirmed when (if) larger size samples become available.

\section{View from the IR: Mg, Ca, and K from APOGEE}

While we were completing the collection of the high-resolution optical dataset,
M\'esz\'aros et al. (2020) published the abundance analysis of 44 GCs from
high-resolution IR spectra acquired in the SDSS-IV APOGEE-2 survey.  The
studied elements include Ca and Mg. Scandium is not present, but they derived
abundances of K, which initially was the only indicator in the literature for the
possible presence of alterations produced in very high temperature H-burning. 
Given the dimension of the dataset and the homogeneous analysis, we decided to
explore their sample to complement the optical observations and support our
previous finding. 

\setcounter{table}{2}
\begin{table*}
\centering
\caption{Results for the APOGEE-2 sample (M\'esz\'aros et al. (2020)}
\begin{tabular}{lrlr|lrlr}
\hline
\hline
GC         &N$_{star}$  & prob &  outliers &NGC         &N$_{star}$  & prob &  outliers  \\
           &            & K-S  &           &           &            & K-S  &              \\
\hline
  0104/47Tuc       & 145 & 0.087                &   0 &  6205/M13         &  28 & 0.244         &   1  \\
  0288             &  38 & 0.190                &   0 &  6218/M12         &  46 & 0.750         &   0  \\
  0362             &  36 & 0.786                &   0 &  6229             &   5 & 0.626         &   0  \\
  1851             &  26 & 0.773                &   0 &  6254/M10         &  60 & 0.479         &   1  \\
  1904/M79         &  20 & 0.168                &   2 &  6388             &   7 & 0.200         &   0  \\
  2808             &  66 & 0.267                &   5 &  6397             &  25 & 0.110         &   2  \\
  3201             &  38 & 0.850                &   0 &  6656/M22         &   6 & 0.900         &   0  \\
  5024/M53         &  12 & 0.169                &   0 &  6715/M54         &   6 & 0.435         &   1  \\
  5139/$\omega$Cen & 263 & 1.6$\times 10^{-4}$  &  34 &  6752             & 108 & 0.047         &   5  \\
  5272/M3          & 133 & 0.741                &   1 &  6809/M55         &  33 & 0.344         &   0  \\
  5904/M5          & 161 & 0.625                &   1 &  6838/M71         &  32 & 0.133         &   1  \\
  6121/M4          & 129 & 0.836                &   0 &  7089/M2          &  21 & 0.379         &   1  \\
  6171/M107        &  42 & 0.211                &   1 &  Pal5             &   5 & 0.375         &   0  \\
\hline
\end{tabular}
\begin{list}{}{}
\item[] N$_{star}$ is the number of stars in each cluster with S/N$\ge$70,
T$_{eff}<5500$ K, and valid [Ca/H] and [Mg/H] abundance ratios.
\end{list}
\label{t:tab3}
\end{table*}

Following M\'esz\'aros et al. (2020), we restricted the APOGEE-2 sample to 26
clusters according to the same selection criteria as they used, excluding
GCs with a higher reddening value ($E(B-V)>0.4$) as well as those with fewer
than five member stars whose spectra have a signal-to-noise ratio $S/N>70$ (see Tab.~\ref{t:tab3} for the
survived clusters).  Almost all of the remaining GCs are already in the optical
sample, so that they can be used as a further check, especially when a larger number
of stars is present, which is often the case. Two GCs are new entries with
respect to the optical samples: NGC~5024 and Pal~5.

In Appendix C we plot the observed distribution of GC stars from the APOGEE-2
sample in the [Ca/H] versus [Mg/H] plane (Fig.~C.1 and Fig.~\ref{f:apo2}). We only
plot and use in the following stars in the range suggested by M\'esz\'aros et
al. (2020), that is, $S/N>70$, T$_{eff}<5500$ K, and actual measurements for Ca and
Mg, excluding stars flagged as limits. No correction for the solar abundances is
applied (the solar reference values are not explicitly provided in M\'esz\'aros
et al.). However, when these selection criteria are applied to the APOGEE
sample,  the differences are small enough that the abundances from IR
spectra nicely fall on the field star distribution. Residual offsets in Ca due
to the different  methods of analysis are taken care of by aligning the mode of
the distributions. 

We then repeated all the steps and statistical tests we performed for the optical
samples. The results are listed in Tab.~\ref{t:tab3} and generally confirm and
strengthen our previous findings.

Three cases need further discussion, however. NGC~2808 and
NGC~6715 (M54) show no statistically significant Ca excess, in contrast with
the optical datasets and the direct visual comparison in Fig.~C.1 and
Fig.~\ref{f:apo2}. On the other hand, the distribution in NGC~6752  appears to
differ from that of field stars according to the K-S statistical test, at
variance with the results from optical  samples. 

The case for NGC~6715 can be easily understood. Only one star out of six
surviving the selection criteria in the APOGEE sample falls outside the
distribution defined by field stars. Neither oxygen nor sodium abundances were
derived for these stars in M\'esz\'aros et al. (2020). However, four
stars are in common in APOGEE and Carretta et al. (2010a), and the outlier star fortunately is one of these. The abundances [O/Fe]$=-0.748$ dex and
[Na/Fe]$=+0.774$ dex explain the classification of this star as an SG star with
extreme alterations in light elements (E component), according to the 
definitions given in Carretta et al. (2009b)\footnote{Primordial P stars are
those with [O/Fe] and [Na/Fe] ratios as in normal halo field stars, the
intermediate I component has depleted [O/Fe] and enhanced [Na/Fe] ratios, and
stars with  [O/Na]$<-0.9$ dex represent the extreme E component.}.  This status
is also confirmed by the high ratio [Al/Fe] found for this star both in
M\'esz\'aros et al. (2020) and Carretta et al. (2010a). It is then not
surprising that its  [Ca/H]$_{\rm corr}$ ratio exceeds 3$\sigma$ (see Section
3), suggesting that the extreme abundances of O, Na, and Al are likely also accompanied by excess of Ca.  On the other hand, all the remaining stars display an
abundance pattern that is perfectly consistent with that of the field stars, and because they
represent $\sim 80\%$ of this limited sample, the statistical test indicates a
similarity with the distribution of field stars, despite the contrasting
evidence from the optical data and the simple visual inspection of the APOGEE
data themselves.

NGC~6752 shows a small skewness in the Ca distribution and the K-S test 
indicates a (quite marginal) difference with the  field star distribution.
Furthermore, a few outliers are present, as also immediately evident from the
visual inspection, and this is not found in the optical samples. We
cross-matched the APOGEE data with ours from the O-Na anti-correlation study by
Carretta et al. (2007),  finding 60 stars in common within the range in S/N and
temperature acceptable for accurate IR data.  All outliers with high
values of Ca are SG stars (as is also the outlier star located below the field
star distribution) according to our classification in P, I, and E components
(Carretta et al. 2009b). For the remaining stars, the distribution of population
is as expected: the vast majority of higher-Ca belongs to SG, while SG and FG
stars are about two-thirds to one-third for the lower-Ca. Finally, we note that
none of the analyses based on optical spectra presented evidence of an excess of
Ca in the sample (Tab.~\ref{t:tab2}). This result is also confirmed by data for
Sc, available for the two optical samples in the literature: the K-S  test excludes
a statistically significant excess of Sc in stars of NGC~6752. In conclusion,
because NGC~6752 shows an Si-Al correlation (Carretta et al. 2009a), but apparently
K is not touched (Mucciarelli et al. 2017) and Ca and Sc are  compatible with
the level found in field stars of similar metallicity in all the available
analyses but the one from APOGEE, the IR data   deserve further
studies.

Finally, the case for NGC~2808 is more complicated; the cluster is a prime
example for extended anti-correlations and the data in our golden set also show
excess in Ca (and Sc, in addition to K, see Mucciarelli et al. 2015 for the
last). The APOGEE data are compatible with the field star distribution, at least
according to the K-S test (see Tab.~\ref{t:tab3}). However, the excess of Ca is
clearly evident from Fig.~C.1, with 5 stars in the sample found at more than
3$\sigma$ above the field star distribution. To better understand this, we
cross-matched the two samples, finding 51 stars in common. By assigning the
common APOGEE stars to the population groups defined in Carretta (2015: P1, P2,
I1, I2, and E,  indicating the unpolluted and increasingly polluted stars), we found
that the analysis by M\'esz\'aros et al. (2020) contains an overabundance of
stars belonging to the FG component (groups P1 and P2), while in Carretta (2015)
these object are shifted in the SG component. More than to selection effects,
this difference is likely to be due to uncertainties in the abundance analysis.
On average, we found that the scatter for the Mg abundances in APOGEE (0.111
dex, 66 stars) is about twice the scatter in the optical analysis (0.056 dex, 83
stars). Larger uncertainties mean that the Ca abundance excess is probably not
perfectly correlated with the population groups, generating  the confusion
signalled in the statistical tests.  The presence of matter processed at
high temperature is well assessed also in the APOGEE sample. In addition to the K-Mg
anti-correlation in NGC~2808 claimed by M\'esz\'aros et al. (2020), confirming
previous results of Mucciarelli et al. (2015), using their APOGEE data we found
a correlation between K and Ca abundances that is significant to a high level
of confidence (p-value$=4.4 \times 10^{-3}$), much more than the weak
anti-correlation between K and Mg (p-value=0.122).

\subsection{NGC~4833: another GC with K-Mg anti-correlation}

The APOGEE data have the additional advantage of providing the K abundance derived
for many stars in several GCs; this is not routinely done. 
Mucciarelli et al. (2012, 2015) found a Mg-K anti-correlation in NGC~2419 and
NGC~2808, respectively, with strongly depleted Mg abundances. M\'esz\'aros et
al. (2020) added $\omega$~Cen to the small list of clusters showing this feature.
They applied very restrictive quality cuts because the two K lines in the H band
are fairly weak, especially at low metallicity, so that confirmation
from optical spectroscopy is required before the extent of the Mg-K
anti-correlation in this third GC can be discussed. Interestingly, M\'esz\'aros et al. (2020) also found a weak anti-correlation in NGC~1904, but they attributed this more 
probably to correlated errors in the measurements.

\begin{figure}
\centering
\includegraphics[scale=0.42]{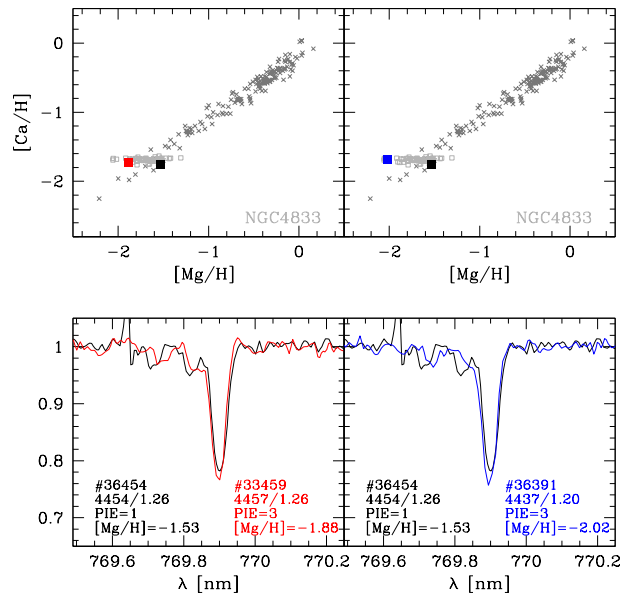}
\caption{Lower panels: Comparison of ESO archive GIRAFFE HR18 spectra in the
potassium resonance K~{\sc i} 7698.98~\AA\ line region for two pairs of stars in
NGC~4833 analysed in Carretta et al. (2014a). Each pair contains two stars
selected to have similar atmospheric parameters (listed in each panel) and to 
span a range as wide as possible in [Mg/H] (also listed). The position of each
pair of stars in the [Ca/H] vs [Mg/H] plane is shown in the upper panels.}
\label{f:kmg4833}
\end{figure}

Unfortunately, the GCs analyzed by M\'esz\'aros et al. (2020) do not include
NGC~4833. The results we obtained here indicate that this cluster as a
good candidate to host a well-defined K-Mg anti-correlation, although it is probably
less extended than those found in NGC~2808 and NGC~2419 (and possibly
$\omega$~Cen). However,we found GIRAFFE spectra with the
HR18 setup in the ESO\ archive (programme 095.D-0539(A), PI Mucciarelli) that were taken to study the K
abundance and its relation to other light elements. In the
same observational programme, spectra in the K~{\sc i} line region were also acquired for stars in NGC~7078 (M~15), NGC~6715 (M~54), and  NGC~5139 ($\omega$
Cen); these are the only GCs that differ in the diagnostic plot presented in
Carretta et al. (2014a,  figure 10) in addition to NGC~4833, NGC~2808, and NGC~2419.
Results from that programme are not yet published, therefore we downloaded spectra of
NGC~4833 from the ESO archive to make a simple test. In Fig.~\ref{f:kmg4833} we
compare the GIRAFFE spectra in the region around the K~{\sc i} 7698.98~\AA\ line
for two pairs of stars selected from Carretta et al. (2014a) to have similar
atmospheric parameters but differences in their Mg content as large as possible.

In both cases the outcome of this comparison is as expected if a K-Mg 
anti-correlation does exist in the stars of NGC~4833: the stars with lower Mg
abundances are also those whose K content is higher (and vice versa), judging from
the line depth. The difference  does not seem very large, but in one case the
chosen stars are not even at the extreme of the Mg range in NGC~4833 (upper left
panel in Fig.~\ref{f:kmg4833}). 

Here we limit ourselves to conclude for the first time that NGC~4833 {\em \textup{is another GC
likely  to host a fully developed K-Mg anti-correlation}}, joining the set of GC
with this signature of extreme nuclear processing in multiple stellar
populations. The K-Mg relation should be explored for the full
sample in NGC~4833, but this is beyond the immediate aim of the present work. Similarly, the analysis of the extant data for NGC~7078 (M~15) would also be
important to ascertain the existence (and extent) of a K-Mg anti-correlation.

\section{GCs with uncertain or borderline results}

As already stated in Sec.~3.4, some GCs in the optical literature
sample  show uncertain results and need further discussion. In some cases,
the uncertainty derives from contradictory indications from different
studies of apparently similar quality and merit. It is not always possible to solve the discrepancies. The solution will probably only be achieve when larger 
samples are studied in a homogeneous way. This will be possible with
high-resolution spectroscopic surveys, such as WEAVE (Dalton et al. 2012) and
4MOST (de Jong et al. 2011), which are soon to start in the Northern and Southern
hemisphere, respectively.

\paragraph{NGC~4590} The cluster has three datasets, only one of which
(Schaeuble et al. 2015) shows GC stars that are statistically different from the field,
according to the K-S test. In this sample, only one star lies more than
3$\sigma$ above the field star distribution, which might be an indication of
processing at high temperatures. Furthermore, we note that Mg seems to
present a wider range of values (see Fig.~A.2) than the other studies. A
possibility is the inclusion in the sample by S15 of blue HB stars, whose
scatter in both Mg and Ca is noticeably higher than RGB and red HB stars. However, the K-S test applied to the [Sc/H] distribution also appears to exclude that
cluster and field stars are extracted from the same parent distribution. 

\paragraph{NGC~6205} The three datasets (two optical, one IR) have similar
size, but only the stars in Cohen \& Melendez (2005) show a possible indication of
high-temperature H-burning from the K-S test and four outliers in Ca. According
to the Sc values, however, there is no significant difference between field and
cluster stars. This GC surely merits further study because it shows very extended
anti-correlations, with an extreme depletion in oxygen.

\paragraph{NGC~6218} For this cluster the two available optical samples contain 
11 stars each, but while nothing is seen in the silver sample dataset, in good
agreement with the results from APOGEE, the K-S test gives contradictory results
for the Johnson \& Pilachowki (2006) case.  However, no star is found to exceed
the 3$\sigma$ threshold. Furthermore, no significant variations in Sc are
evident from the comparison with field stars (Fig.~\ref{f:B5}).  We have no
explanation for this, and while the GC probably does not show Ca enhancement,
larger samples are required to  firmly settle the issue. 

\paragraph{NGC~6440} Only seven stars are present in Mu\~noz et al. (2018) for
this GC, for which the statistical test indicates a difference with the field
star distribution. NGC~6440 is metal rich, at the limit of the Gratton et al.
(2003) sample. However, the situation remains the same when we compare it to more
extended samples of field stars (e.g. Bensby et al. 2014). When we
instead consider Sc, the cluster distribution does not statistically differ from the
field stars. We also note that the outlier farther away from the field
distribution has a ratio [Na/Fe]=+0.15 dex that identifies it as an FG star,
unlikely to have severe alterations in any of the proton-capture elements. Given
the contradictory results of Ca and Sc and the small sample, further tests are
required. 

\paragraph{NGC~6441} Abundances for this cluster come from Gratton et al. (2006,
2007). At variance with NGC~6388, which is often considered a sibling, it apparently
shows indications of hot-H cycling, with a statistically significant excess in Ca
according to the K-S test, and with many outliers, while Sc seems to be
the same as in the field star distribution. The comparison with the
analysis of NGC~6388, an almost twin cluster in metallicity, age, and location
in the Galactic bulge, may be of some help. The observed spreads in [Mg/H],
[Ca/H], and [Sc/H] are 0.06, 0.06, and 0.08 dex, respectively, for NGC~6388
(Carretta and Bragaglia 2020). These values increase to 0.18, 0.24, and 0.21 dex
for NGC~6441 in Gratton et al. (2006, 2007). However, we remark that only 34\%
of the requested observations were obtained for NGC~6441, while observations
were completed for NGC~6388 (and are further complemented by many archival
datasets in the forthcoming analysis we used here, Carretta \&  Bragaglia,
in preparation).  As a consequence, the median S/N for HR13 spectra in NGC~6441
is 55, compared to 76 for NGC~6388. We therefore conclude that the larger
dispersions observed in NGC~6441 are probably not intrinsic, but are mostly  due
to the lower quality of the spectra. A definitive answer must unfortunately
await until better quality data are available also for NGC~6441.

\paragraph{NGC~6553} For this metal-rich GC, the dataset by Tang et al. (2017)
shows no difference from the field stars, while the  dataset by Johnson et al.
(2014), similar in size, does (even if the K-S test is only marginally
conclusive). However, when we consider the enhanced set of metal-rich field stars
(Bensby et al. 2014), the evidence weakens as the K-S test is passed. Therefore NGC~6553 is probably not a strong enough case for Ca enhancement.

\paragraph{Pal~14} For this very distant GC, the Ca and Sc distributions give
contradictory answers. Given the rather large spread in Mg and because the
outliers are below the field star distribution and not above, we tend to
attribute the result to a combination of small sample and large errors.

\section{Discussion and summary}

Star-to-star variations in the light proton-capture elements C, N, O, and Na are so
ubiquitous in Milky Way GCs that the presence of the O-Na anti-correlation has
been proposed to be the distinctive definition of a genuine GC
(Carretta et al. 2010c) because the overwhelming majority of GCs show this
feature (see the reviews by Bastian \& Lardo 2018, Gratton et al. 2019).  Only a
subset of GCs, the most massive and/or metal-poor, are known to host
anti-correlated variations in the abundances of the higher mass Mg and Al (Carretta
et al. 2009a, Pancino et al. 2017, Nataf et al. 2019, M\'esz\'aros  et al.
2020). The present census searched for indications of much higher
temperatures reached in the FG polluters of multiple population in GCs, in a
regime where the production of Al was disfavoured and an excess of K, 
Ca, and Sc were instead likely produced. Recalling that some GCs had controversial
results according to the test we adopted, we focus in the following on the
eight GCs with robust detections, those where the statistical criteria all
converged at a high level of confidence.

\subsection{Excesses and multiple populations}

In Tab.~\ref{t:tab4} we summarise the GCs and  studies in which variations in K,
Ca, and Sc abundances, related to alterations in the Mg amount, were detected,
confirmed,  or simply noted. Significant variations in Ca with respect to the
level of field stars are found in all these GCs, whereas K and Sc abundance
variations are uncovered only in about half of the sample. 

Indications for such variations have been reported in a few of the original
papers. In NGC~4833, evidence of a Sc-Ca correlation was detected by Carretta
et al. (2014a), but it was deemed to be scarcely significant given the large internal
errors involved. This was consistent with burning of Mg under conditions
advocated by Ventura et al. (2012) to explain the K-Mg anti-correlation in
NGC~2419, however.  Again for NGC~4833, Roederer and Thompson (2015) noted that [K/Fe]
appeared to be anti-correlated with [Mg/Fe], but they found the difference in K
abundances in the Mg-rich and Mg-poor groups significant only at about
1.5$\sigma$. However, when we perform a linear regression on their limited sample
(excluding star 4-224 with [K/Fe]=+1.39, as they did), we find a p-value=0.018,
which is statistically significant.  A large spread in [Mg/Fe] was detected in
NGC~5824 by Mucciarelli et al. (2018) and Roederer et al. (2016). The
latter study did not detect evidence of  internal spread for other
$\alpha-$element ratios. Johnson et al. (2017a) did not measure K and Sc
abundances in NGC~5986, but they clearly stated that a wide range in [Ca/Mg]
was found in this cluster. Although weaker than the spreads in NGC~2808 or
NGC~2419, they concluded that significant processing at high temperature was
probably at work in this cluster as well, as implied by the diagnostic plot [Ca/Mg]
versus [Ca/H] shown in Carretta et al. (2014a).  Finally, in their analysis of
NGC~6402, Johnson et al. (2019) detected a correlation  between [Al/Fe] and  [Ca/Fe], which may indicate possible Ca variations.
However, they also noted the need for similar correlations involving K and/or Sc
to confirm the high-temperature nuclear processing.

\setcounter{table}{3}
\begin{table*}
\centering
\caption{Globular clusters with detected excesses in K, Ca, and Sc.}
\begin{tabular}{llll}
\hline
\hline
NGC     & K                          & Ca                         &Sc   \\
\hline
  2419  & -Cohen \& Kirby (2012)      & -Cohen \& Kirby (2012)      & -Cohen \& Kirby (2012)   \\
        & -Mucciarelli et al. (2012)  & -Mucciarelli et al. (2012)  & -Carretta et al. (2013a)   \\
        & -Carretta et al. (2013a)    & -Carretta et al. (2013a)    & -this work   \\
        &                             & -this work                  &    \\
        &                             &                             &    \\
  2808  & -Mucciarelli et al. (2015)  & -Carretta (2015)            & -Carretta (2015)   \\
        & -M\'esz\'aros et al. (2020) & -Carretta et al. (2014a)    & -this work   \\
        &                             & -this work                  &    \\
        &                             &                             &    \\
  4833  & -Roederer \& Thompson (2015)& -Carretta et al. (2014a)    & -Carretta et al. (2014a)   \\
        & -this work                  & -this work                  & -this work   \\
        &                             &                             &    \\
  5139  & -M\'esz\'aros et al. (2020) & -Carretta et al. (2013a)    & -this work   \\
        &                             & -Carretta et al. (2014a)    &    \\
        &                             & -this work                  &    \\
        &                             &                             &    \\
  5824  &                             & -this work                  &    \\
        &                             &                             &    \\
  5986  &                             & -Johnson et al. (2017a)     &    \\
        &                             & -this work                  &    \\
        &                             &                             &    \\
  6402  &                             & -Johnson et al. (2019)      &    \\
        &                             & -this work                  &    \\
        &                             &                             &    \\
  6715  & -M\'esz\'aros et al. (2020) & -Carretta et al. (2013a)    &    \\
        &                             & -Carretta et al. (2014a)    &    \\
        &                             & -this work                  &    \\
\hline
\end{tabular}
\label{t:tab4}
\end{table*}

The novelty and effectiveness of the approach adopted in the present work also 
rests on the unambiguous confirmation of all the previous indications of suggested
detections. To verify that the variations in Ca (as well as in Sc, whenever
possible) are   produced by the multiple-population phenomenon, we also
mapped in  every GC the stars with Ca and Mg abundances onto the O-Na
anti-correlation, the main signature of multiple populations in GCs. We used the
P, I, and E classification defined in Carretta et al. (2009a; see also Section
4, this work), except for NGC~2808 (Carretta 2015), NGC~5986 (Johnson et al.
2017a) and NGC~6404 (Johnson et al. 2019), for which we followed the  subdivision
into components given in the original studies.

\begin{figure*}
\centering
\includegraphics[scale=0.75]{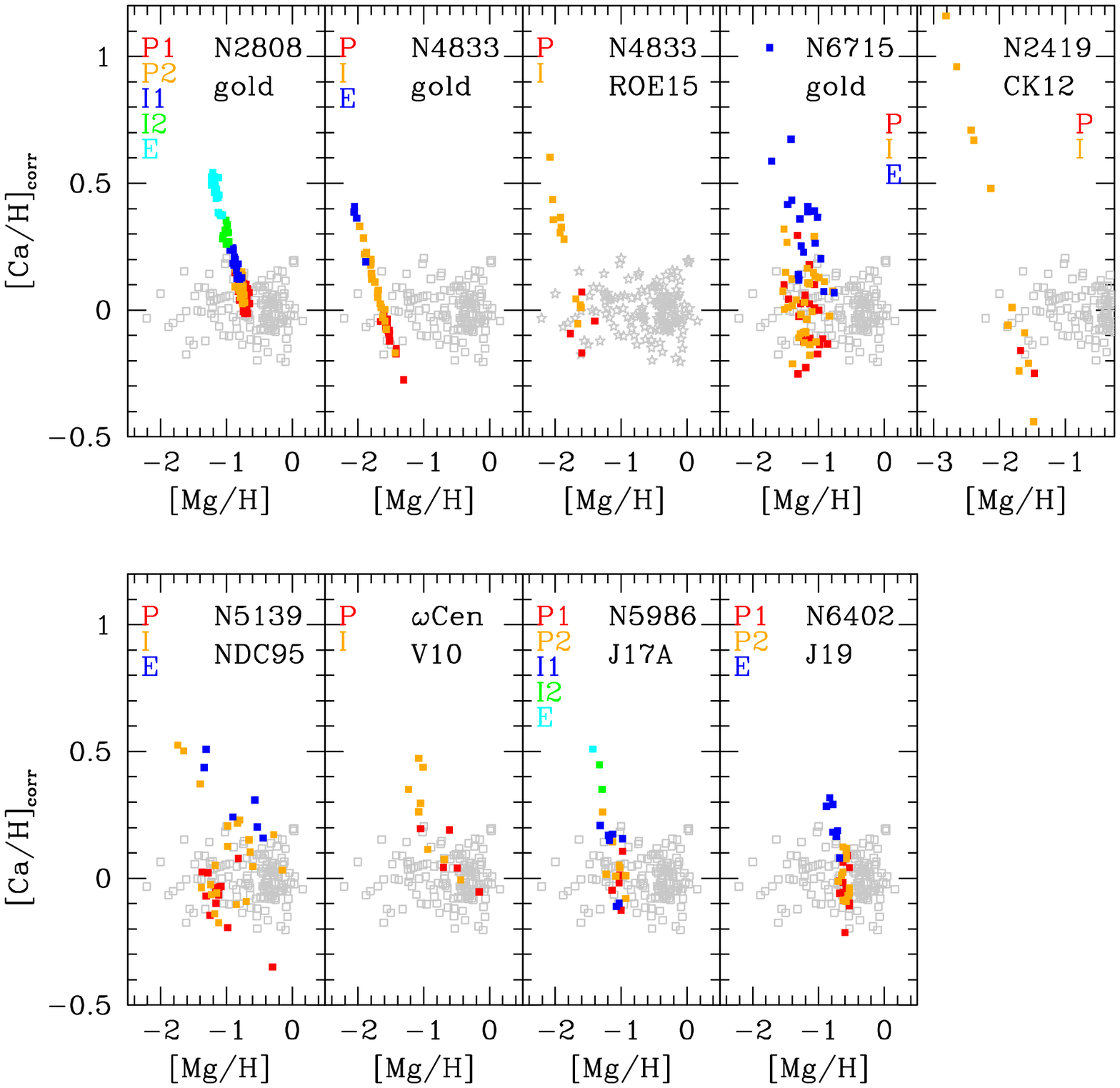}  
\caption{Linearised ratios [Ca/H]$_{\rm corr}$ as a function of [Mg/H] ratios
for GCs with robust excess detections (this work). Cluster stars are coloured
according to whether they belong to the P, I, or E components defined by the O and Na
abundances. For NGC~2808 (Carretta 2015), NGC~5986 (Johnson et al. 2017a) and
NGC~6404 (Johnson et al. 2019), we followed the original  division into
components. }
\label{f:PIEokca}
\end{figure*}

The association between Ca excesses (as detected by the linearised  [Ca/H]$_{\rm
corr}$ and the P, I, or E classification is summarised in Fig.~\ref{f:PIEokca} for
seven GCs. In this plot, only stars with  derived abundances of light elements
(O and Na) required for P,I, and E classification are shown. For NGC~5824, O and Na
abundances are available only for two stars with MIKE spectra, so that it is not
possible to separate the sample into P, I, and E groups in this cluster. For NGC~2419
(CK12)  only three stars have O abundances; for this GC we can only distinguish
between first- (i.e. P) and second-generation (i.e. I) stars. In the majority of
cases, the stars showing the highest Ca excess fall in the E component or among
the most polluted I stars, where the composition changes due to a previous
stellar generation were driven towards more extreme Na enhancements and O
depletions. This figure is proof that in these GCs the changes in light
proton-capture elements such as O and Na, as well as the alterations in Ca (and by
extension, in K and Sc, as in NGC~2808), are linked by the same mechanisms. 

To better quantify this association, we made linear fits to the observed data,
using the [Ca/H] ratio as independent variable. We found statistically
significant correlations and/or anti-correlations with [Mg/H] in NGC~2419
(p=0.003); with [O/H] (p=0.040), [Na/H] (p=4$\times 10^{-4})$, and [Al/H] 
(p=0.017) in NGC~2808; with [Na/H] (p=0.033), [Si/H] (p=0.004), and [K/H]
(p=0.009) in  NGC~4833 (ROE15); with [Na/H] and [Al/H] (p$<1\times 10^{-6}$) in
NGC~5139 (Ndc95);  with [Na/H] (p=0.011) and [Al/H] (p=3$\times 10^{-4}$ in
NGC~5986 (J17A); with [Na/H]  (p=0.001) and [Al/H] (p=0.008) in NGC~6402 (J19);
and with [Na/H] (p$<1\times 10^{-6}$) in  NGC~6715 (Carretta et al. 2010a).
The tight association between light and higher mass proton-capture elements  then
indicates that this phenomenon is seen in GCs where the production of nuclearly
processed matter in the polluters, regardless of they were, did proceed more than in
other clusters, resulting in extreme modifications in stellar composition.

\begin{figure}
\centering
\includegraphics[bb=18 420 592 718,clip,scale=0.45]{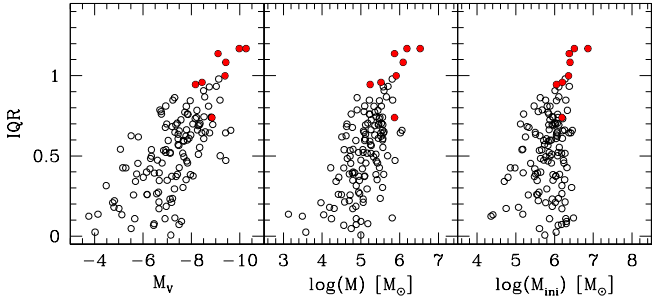}
\caption{Extension of the Na-O anti-correlation as given by the empirically
calibrated IQR2[O/Na] values from Carretta (2019) as a function of total
absolute magnitude $M_V$ from Harris (1996, 2010 on-line edition; left panel)
and of the present-day and original total GC masses (middle and right panels,
respectively) from Baumgardt et al. (2018). GCs with robust detection of changes
in Ca (K and Sc) in the present census are indicated with filled red circles.}
\label{f:iqr2ok}
\end{figure}

Carretta (2006) introduced a parameter that is well suited to quantify the range of
abundance ratios reached in GC multiple population: the interquartile range of
the [O/Na] ratio. The higher IQR[O/Na], the lower the O and the higher the Na levels observed in a GC. The extent of the O-Na anti-correlation is found to be
primarily driven by the total GC mass, as indicated by the tight correlation
between IQR[O/Na] and the total absolute cluster magnitude $M_V$ (Carretta et
al. 2010c). Because we have a truly homogeneous analysis only for GCs in our
golden and silver  samples, we adopted the  IQR[O/Na] values (IQR2) from
the recent empirical calibration by Carretta (2019). Using homogeneous O and Na
values for 22 GCs, an empirical calibration as a function of $M_V$, cluster
concentration, and HB index was given for non-post-core-collapse GCs, providing
empirical IQR2s on the same scale for 95 GCs in the Milky Way. These
homogeneous IQR2s are available for most GCs in the present
census\footnote{NGC~5139 does not have an HB index, therefore we simply
adopted the observed IQR[O/Na] value from data by Johnson and Pilachowski 
(2010) for this cluster.} , and they are plotted in Fig.~\ref{f:iqr2ok} as a function of the
cluster masses, using the total absolute magnitude $M_V$ (as a proxy for total
mass) and both the present-day and the initial mass from Baumgardt et al. (2018).

The eight GCs with robust detection of variations in Ca have the highest extent
of the O-Na anti-correlation in the Galaxy, with the exception of NGC~5824, and
are among the most massive GCs in the Milky Way. The processes of mass loss  do
not seem to have affected the location of these GCs in the plane IQR2-mass strongly,
except for a small increase in the present-day mass spread. All the GCs with a
high probability of hosting internal abundance variations in Ca (K, Sc) are then at
the high-mass end of the Galactic distribution. However, some
other factor(s) must be involved, probably related to the cluster metallicity. Although NGC 6388 is one of the most massive GCs in the Milky Way, it does not show any
significant change in the content of these higher mass proton-capture elements, and
we note that its metallicity is about 0.6 dex higher than that of NGC~2808, the
cluster with the highest [Fe/H] values of the GCs highlighted in
Fig.~\ref{f:iqr2ok}.

The metallicity effect is also evident from the fact that all these eight
GCs belong to the OoII groups, according to the mean period of their RR Lyrae
stars, with the exception of the two most metal-rich clusters, NGC~2808 and
NGC~6402 (e.g. Catelan 2009). All of them are classified as old halo GCs and
are associated with present-day disrupting dwarf galaxies or with inferred past
accretion events (see Massari et al. 2019, Forbes 2020, Kruijssen et al. 2020).
This origin does not appear to be a strong constraint because orbital properties
and age-metallicity relations appear to indicate that most GCs entered the main
Galaxy following  the accretion of their once parent smaller galaxies. The nomenclature of the ancestral systems almost alone varies in different studies.

\begin{figure}
\centering
\includegraphics[bb=18 180 420 500,clip,scale=0.45]{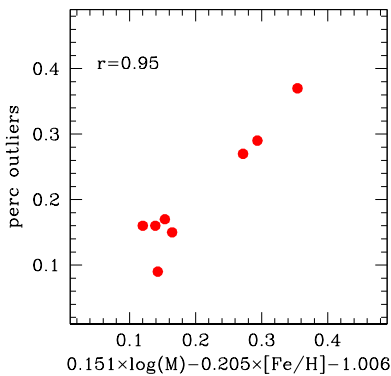}
\caption{Relation between mass and metallicity and the fraction of outliers in
[Ca/H]$_{corr}$ in the eight clusters with robust detection.  The multivariate
relation is indicated in the figure x-axis; the Pearson $r$ coefficient is
shown.}
\label{f:biv}
\end{figure}

To better explore the problem, we computed the fraction of outliers (i.e. GC
stars with high and very high values of Ca with respect to the distribution of
field stars) in the eight GCs. For NGC~2419 and NGC~5139, where the different
analyses provide similar fractions that agree excellently, we averaged them; for
NGC~4833, where differences are large, we opted for our analysis from the golden
sample, based on a sample three times larger than in ROE15. We checked whether
these fractions correlate with many global cluster parameters. We found that
good correlations exist with mass and metallicity (from Baumgardt et al. 2018, 
and the Harris 1996, 2010 on-line edition, respectively). The best correlation 
is a multivariate relation where the fraction of stars with more altered Ca
excesses is a function of both parameters, shown in  Fig.~\ref{f:biv}. The
Pearson correlation coefficient is $r_P=0.95,$ which for eight objects implies
a very high statistical confidence level for the derived relation
($p=3\times10^{-4}$), corroborating the reality of this new observable we
introduced in the context of multiple populations in GCs. This type of
dependence on mass and [Fe/H] has been found previously for other manifestations of
the multiple population phenomenon in GCs, for example for the minimum level of O
reached in GCs (Carretta et al. 2009b) and the maximum value of Al production
(Carretta et al. 2009a). Altogether, these observations  reinforce the findings
of Carretta et al. (2010c), who showed that the multiple population signatures
are driven by a few cluster parameters and that mass and metallicity are among
the most important ones.

Eight out of a sample of 77 GCs is a 10\% fraction of clusters showing
significant variations in high-mass proton-capture elements. In the same subset, a
rough estimate is that the Mg-Al anti-correlation is observed in 30 GCs, that is,
39\% of the sample.  The source of the variations surveyed in the present work
seems to be $\text{rare}$ (and dependent on mass and metallicity), and some sort of
on-off mechanism is apparently  required.

Two modalities are possible: either the polluters are normal stars
sampled from the normal IMF of GCs and this particular processing occurred only
under particular conditions, or some type of polluter only appeared in a few
GCs for some reasons. The best candidates for the first hypothesis could be
massive AGB and super-AGB stars because at very low metallicities their hot-bottom burning may reach temperatures so high that nucleosynthesis may
advance up to high-mass species such as K (Ventura et al. 2012). A problem might
then arise for the higher metallicity GCs such as NGC~2808 and NGC~6402 because the
mechanism was used to explain the chemical pattern in NGC~2419, with [Fe/H]$\sim
-2$ dex.

In the second case, a plausible on-off mechanism might be represented by the
supermassive stars born at the centre of some GCs.  Gieles et al. (2018) showed
that the required conditions for these objects to form may be limited to the
most massive clusters ($\gtrsim 10^6$ stars) with high accretion rates ($\gtrsim
10^5$ M$_\odot$/Myr). The first condition is clearly satisfied by the above
eight GCs, which are among the most massive in the Galaxy, and with scarce evidence for a
significant mass loss. Denissenkov and Hartwick (2014) showed that  such  stars 
might  produce  gas  whose content matches  the composition  of  E  stars.
However, the synthesis was followed only up to Mg and Al abundances, nothing was
said about higher mass elements.

On the theoretical side, it would be interesting to have a self-consistent and
detailed  pattern of nucleosynthesis including all species involved in the
phenomenon of multiple population for the whole metallicity range spanned by
GCs. Now we know that changes in the chemical content of SG stars may concern a
range of species that is wider than previously thought, at least in some GCs. 

For the observers, we need to ascertain the existence of these excesses also in GCs with uncertain evidence, and we need to gather better
statistics for each GC in general. An effort to analyse all the GIRAFFE spectra for GCs in
our silver sample is already in progress with the aim to move them to the golden sample. This would increase the number of stars with homogeneous
abundances of Mg, Ca, and Sc derived in each cluster by about one order of magnitude.

\subsection{Summary}

We have extended the analysis presented in Carretta and
Bragaglia (2019) of the two GCs NGC~2808 and NGC~6388 to all available data of
GCs where information on Ca (an possibly Sc) is available. We collected useful
data  for 77 individual clusters, which is a significant fraction of the MW
globulars. Our prime sample contains 10 GCs that were homogeneously analysed by our group
and  with large number of stars observed with GIRAFFE and UVES, followed by 16
GCs where only UVES spectra are available. To this we added literature samples, 
both from optical and IR spectra (the latter from the homogeneous APOGEE
survey). For many clusters,  more than one source is available.

We verified whether overabundances of Ca (and Sc) over the normal field stars
level are present in these GCs, using the diagnostic plot [Ca/H] versus [Mg/H].
We then quantified the enhancement using a K-S test and computing the fraction 
of the outliers (at 3$\sigma$ level) in each GCs with respect to the field star
distribution. We found  a clear indication of Ca enhancement (i.e. of very high
temperature H-burning) in NGC~4833, NGC~6715, NGC~6402, NGC~5296, NGC~5824, and
$\omega$~Cen, which join the previously studied NGC~2419 and NGC~2808 (Cohen and
Kirby 2012, Mucciarelli et al. 2012; Carretta 2015, Carretta and Bragaglia 2019,
respectively). For these clusters we connected the Ca enhancement with the
normal variations in light elements (O, Na), which are ubiquitous in GCs. We
found that the higher levels of Ca are associated with the more polluted SG stars that
belong to the E and more extreme I populations.  

A few more clusters are uncertain cases either because too few stars for robust
statistical analysis are available, or because different datasets give
contradictory results (or both). We expect that the situation becomes
clearer and a better census is possible when a) the full analysis of the 
GIRAFFE spectra of the 16 GCs in our silver sample is completed, which will increase the number of stars tenfold, and b) in the longer run, when data
from large spectroscopic surveys such as WEAVE and 4MOST become available.

The eight GCs with a robust detection of Ca (and possibly Sc) enhancement are
among the most massive in the MW, as measured both from present-day and initial
mass. They also have very extended O-Na anti-correlations, as measured by 
IQR[O/Na], strengthening the relation between Ca excess and multiple populations
in GCs. The fraction of outliers in Ca abundance also
correlates well with cluster mass and metallicity. It would be very useful if models
for polluters could systematically explore the yields up to at least Ca over
different masses and metallicities.

Finally, we suggest for the first time that an anti-correlation between Mg and K
also exists in NGC~4833 (in addition to the three known cases of NGC~2419, NGC~2808, and
$\omega$~Cen; Cohen and Kirby 2012, Mucciarelli et al. 2012, M\'esz\'aros et al.
2020). We hope that the case will be studied in detail, together with other good
candidates such as NGC~7078, NGC~6715 (see Carretta et al. 2014a), NGC~1904
(M\'esz\'aros et al. 2020), and all other GCs for which we found Ca enhancements
in the present work.

\begin{acknowledgements}
We thank D. Romano, L. Moscardini, P. Montegriffo, M. Bellazzini, and R.
Schiavon for useful comments. We thank the referee for their careful reading and
useful discussion of the manuscript. This research has made use of the SIMBAD
database (Wenger et al. 2000), operated at CDS, Strasbourg, France and of the
VizieR catalogue access tool, CDS, Strasbourg, France (DOI:
10.26093/cds/vizier). The original description of the VizieR service was
published in Ochsenbein et al. 2000). This research has made use of NASA's
Astrophysics Data System. We made extensive use of {\sc TOPCAT}
(http://www.starlink.ac.uk/topcat/, Taylor 2005).
\end{acknowledgements}

\clearpage

\begin{appendix}
\section{Observed distributions for the literature samples}

In this appendix we show a visual catalogue of the literature samples examined
in the present work that are based on optical high-resolution spectroscopy (Figures
A1-A10).

\begin{figure*}[h]
\centering
\includegraphics[scale=0.90]{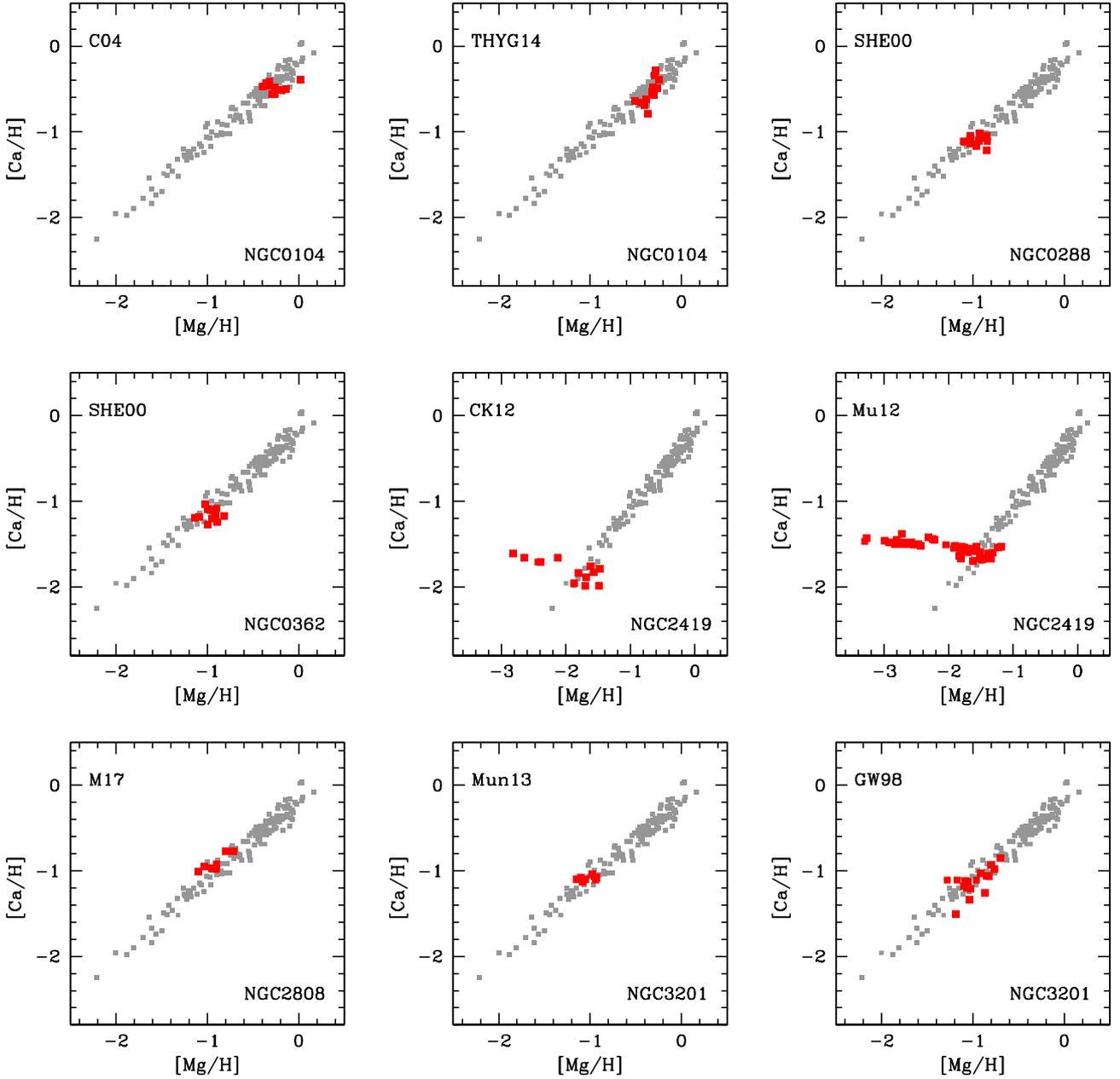} 
\caption{Observed [Ca/H] ratios as a function of [Mg/H] ratios for nine GCs  in
the literature sample, ordered by NGC name. In each panel we list the alphanumeric code
of Tab~\ref{t:tab1} to identify the corresponding study. The [Mg/H]
scale for NGC~2419 is different, to encompass the whole range of variation in
this cluster. The GC stars are plotted in blue whenever the sample includes fewer
than five objects.}
\label{f:A1}
\end{figure*}

\begin{figure*}[h]
\centering
\includegraphics[scale=0.90]{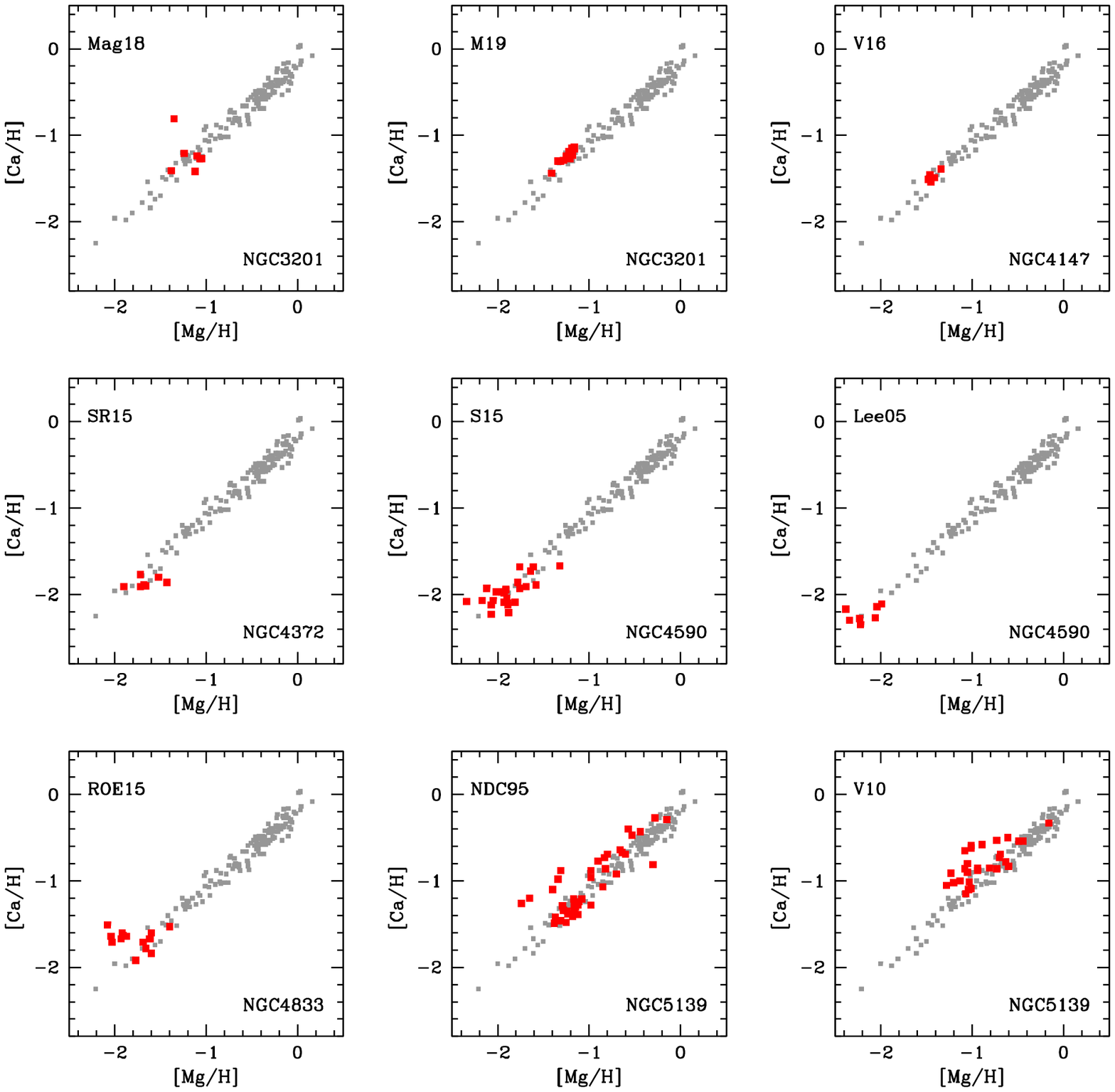} 
\caption{As in Fig.~\ref{f:A1} for the other GCs in the literature sample.}
\label{f:A2}
\end{figure*}

\begin{figure*}[h]
\centering
\includegraphics[scale=0.90]{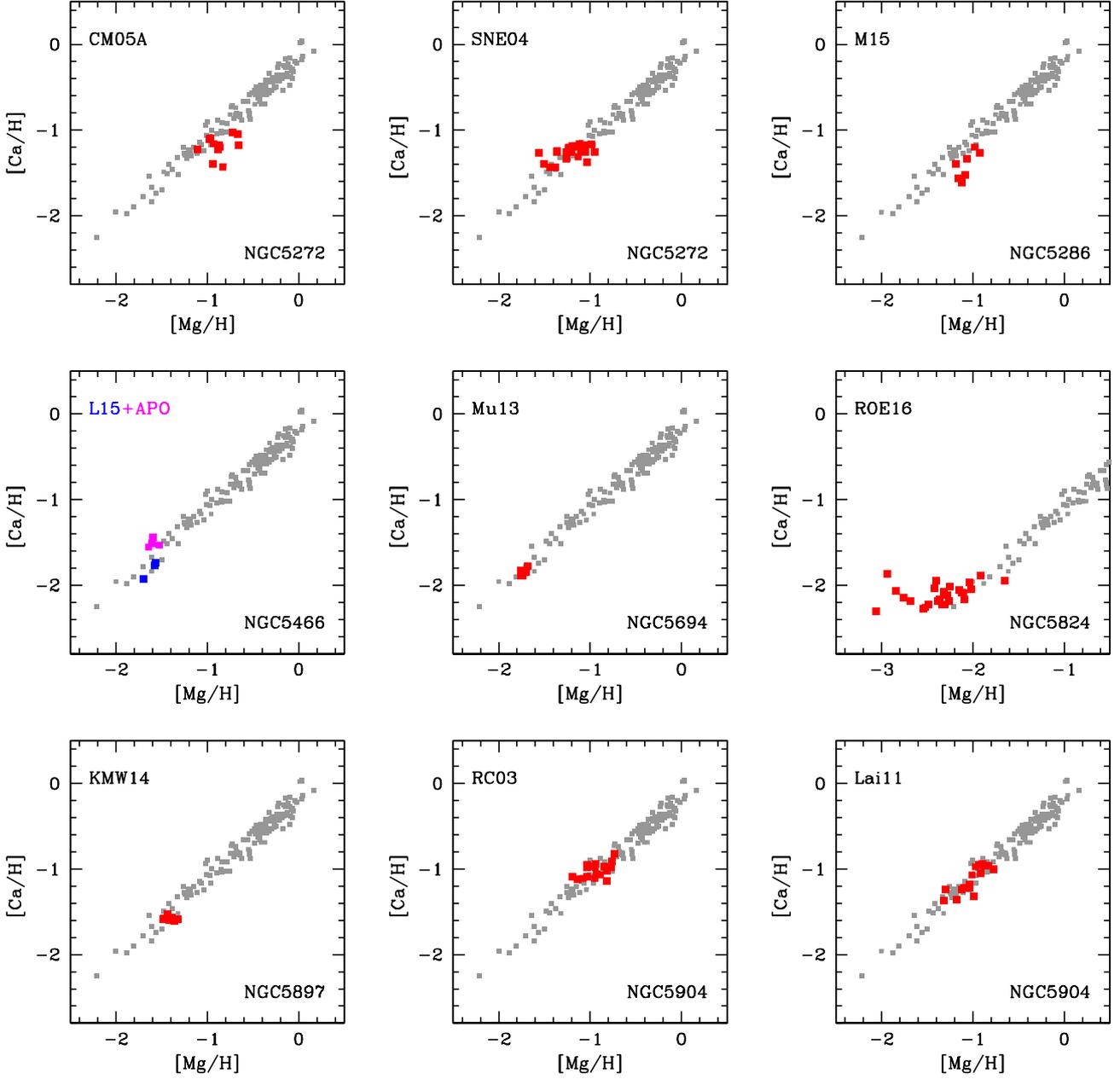} 
\caption{As in Fig.~\ref{f:A1} for the other GCs in the literature sample. For
NGC~5466 we also show the small APOGEE sample by M\'esz\'aros et al. (2020), in
magenta.}
\label{f:A3}
\end{figure*}

\begin{figure*}[h]
\centering
\includegraphics[scale=0.90]{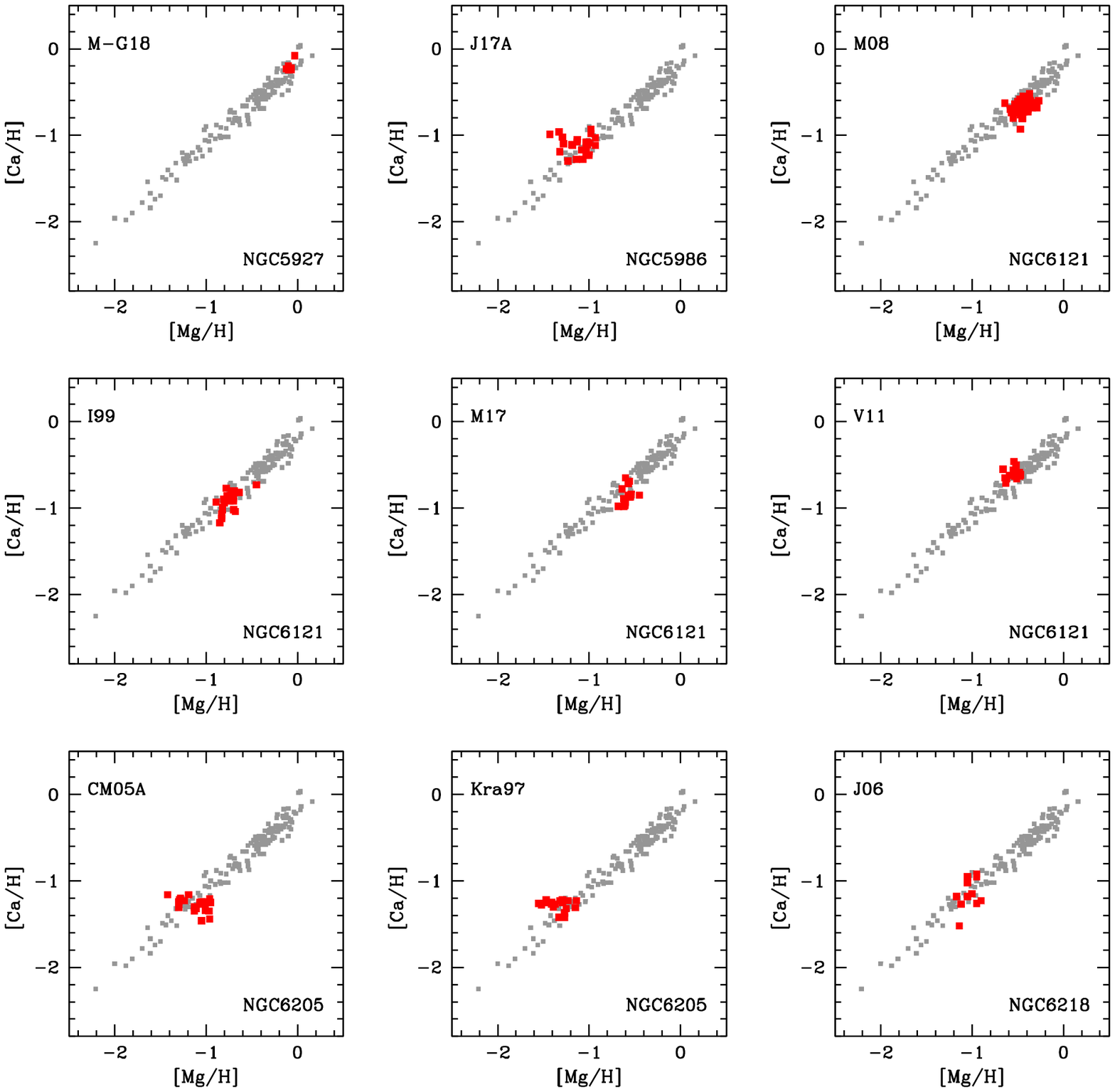} 
\caption{As in Fig.~\ref{f:A1} for the other GCs in the literature sample.}
\label{f:A4}
\end{figure*}

\begin{figure*}[h]
\centering
\includegraphics[scale=0.90]{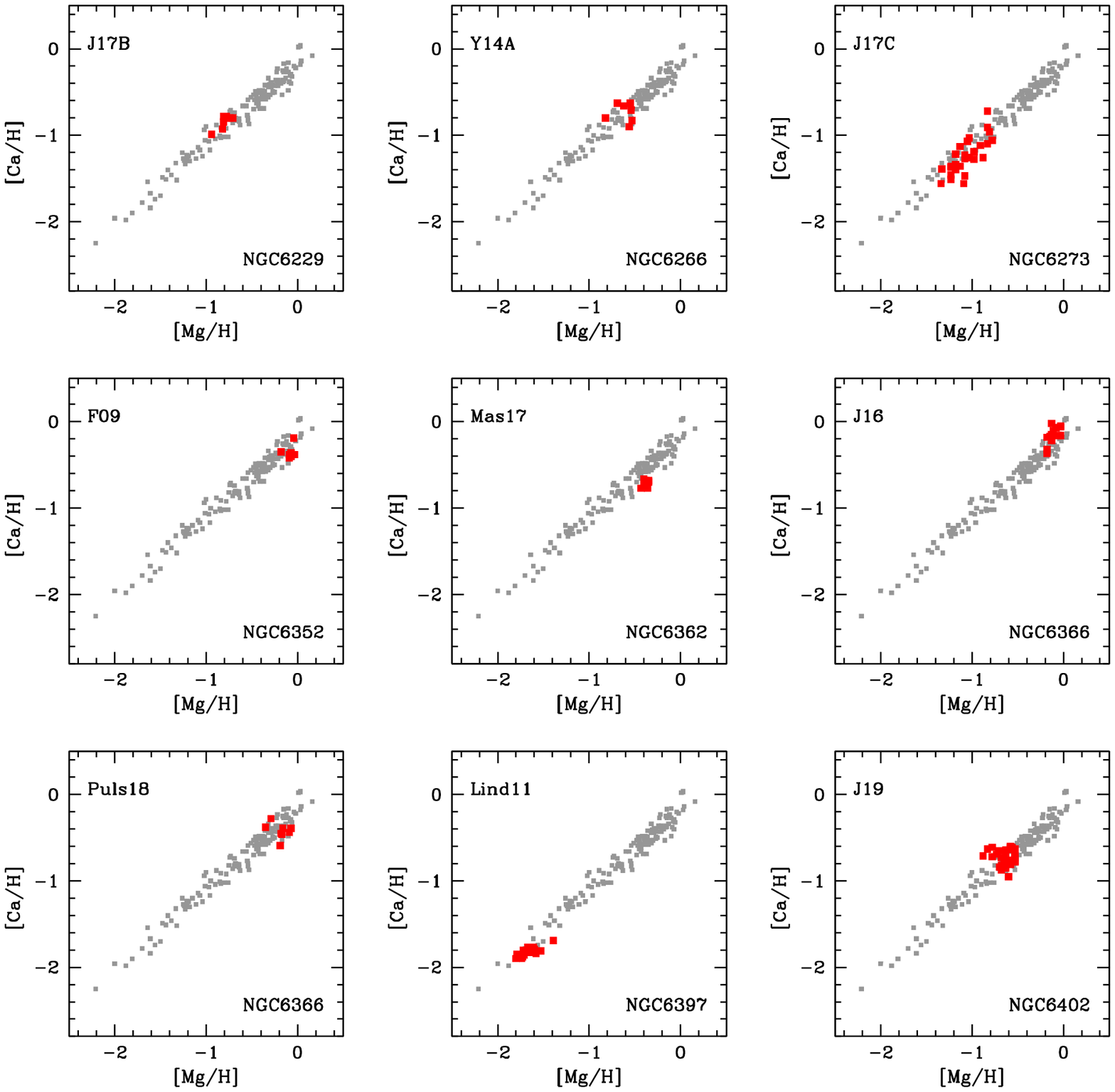} 
\caption{As in Fig.~\ref{f:A1} for the other GCs in the literature sample.}
\label{f:A5}
\end{figure*}

\begin{figure*}[h]
\centering
\includegraphics[scale=0.90]{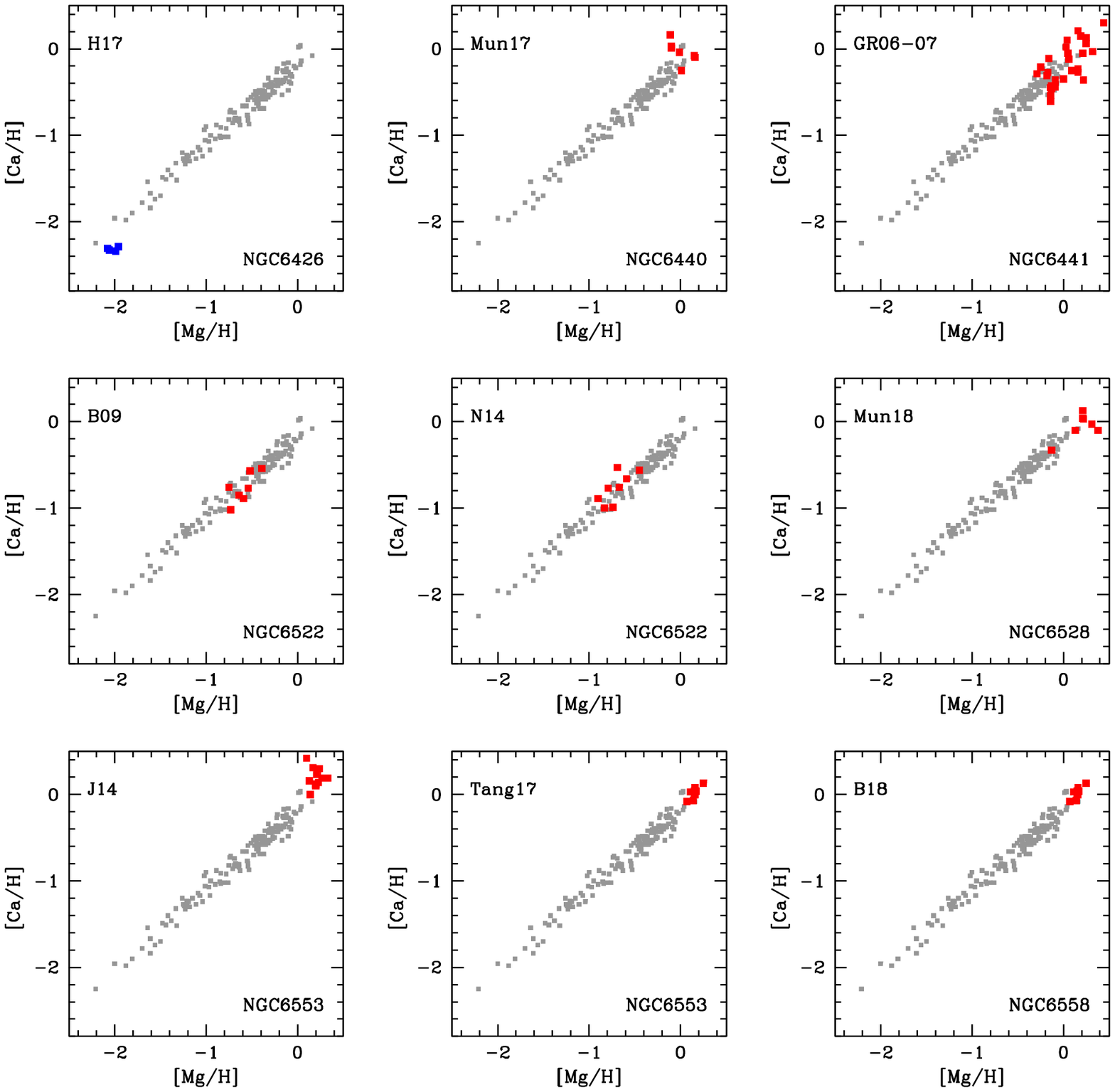} 
\caption{As in Fig.~\ref{f:A1} for the other GCs in the literature sample.}
\label{f:A6}
\end{figure*}

\begin{figure*}[h]
\centering
\includegraphics[scale=0.90]{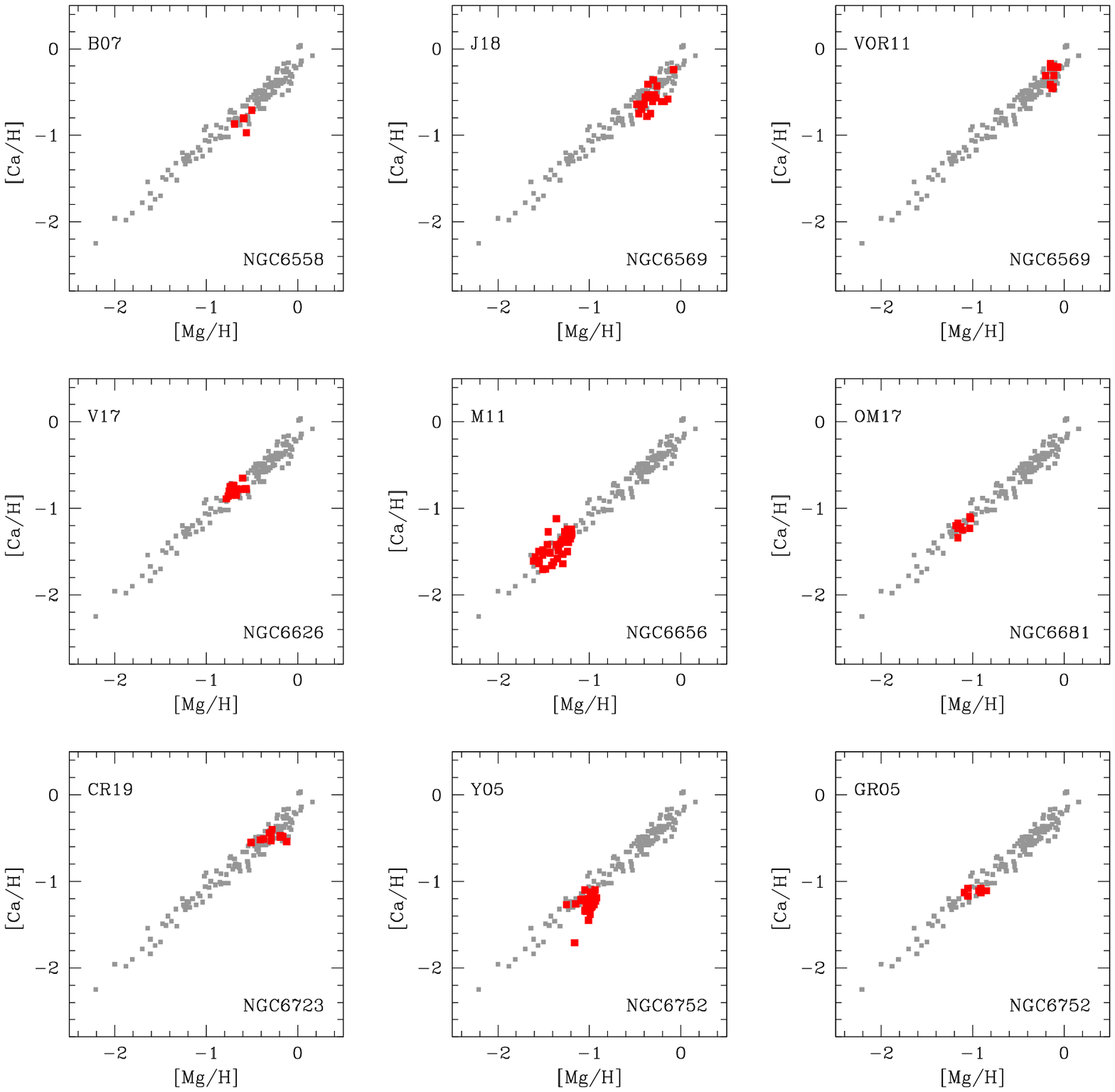} 
\caption{As in Fig.~\ref{f:A1} for the other GCs in the literature sample.}
\label{f:A7}
\end{figure*}

\begin{figure*}[h]
\centering
\includegraphics[scale=0.90]{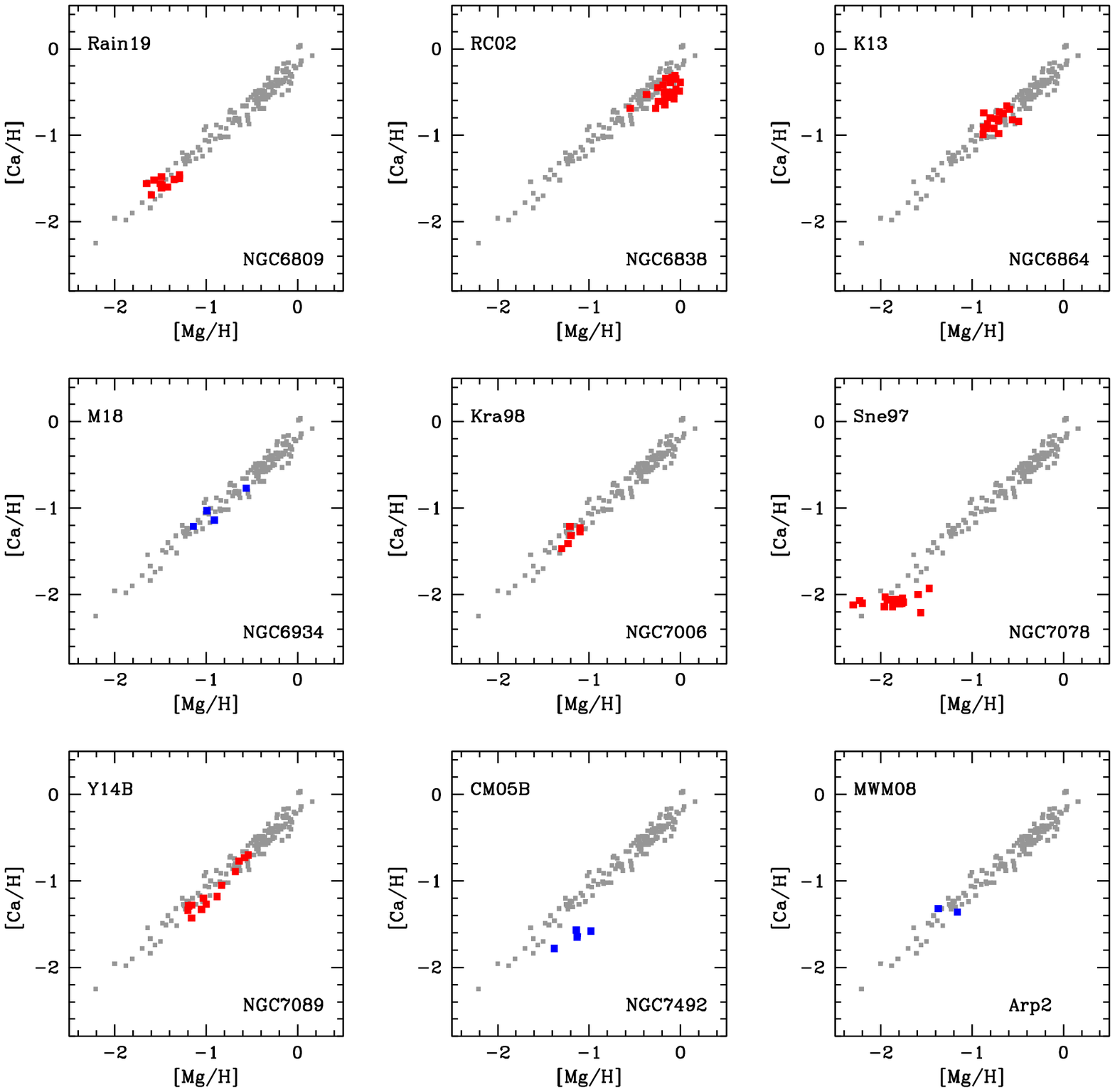} 
\caption{As in Fig.~\ref{f:A1} for the other GCs in the literature sample.}
\label{f:A8}
\end{figure*}

\begin{figure*}[h]
\centering
\includegraphics[scale=0.90]{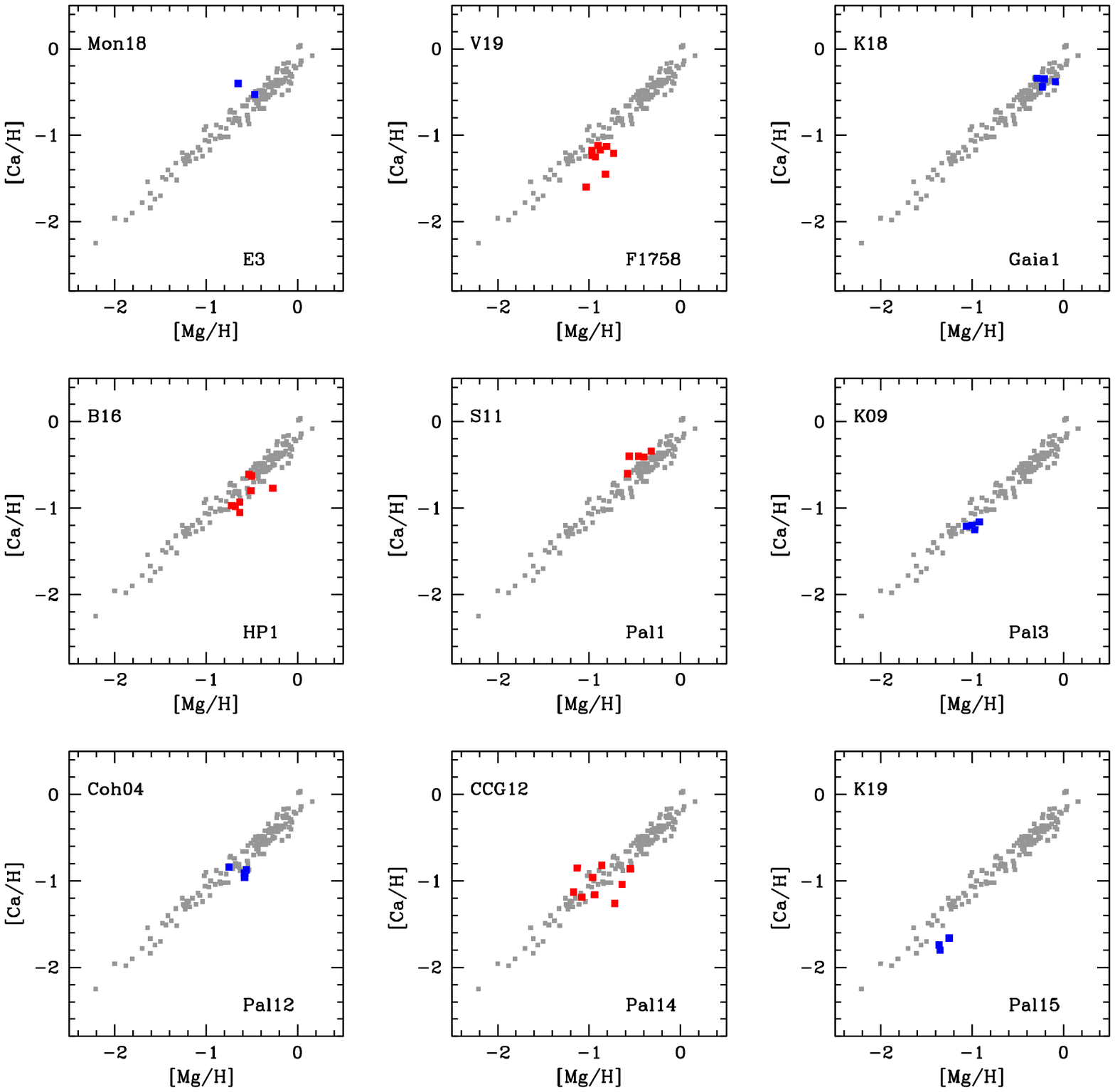} 
\caption{As in Fig.~\ref{f:A1} for the other GCs in the literature sample.}
\label{f:A9}
\end{figure*}

\begin{figure*}[h]
\centering
\includegraphics[scale=0.90]{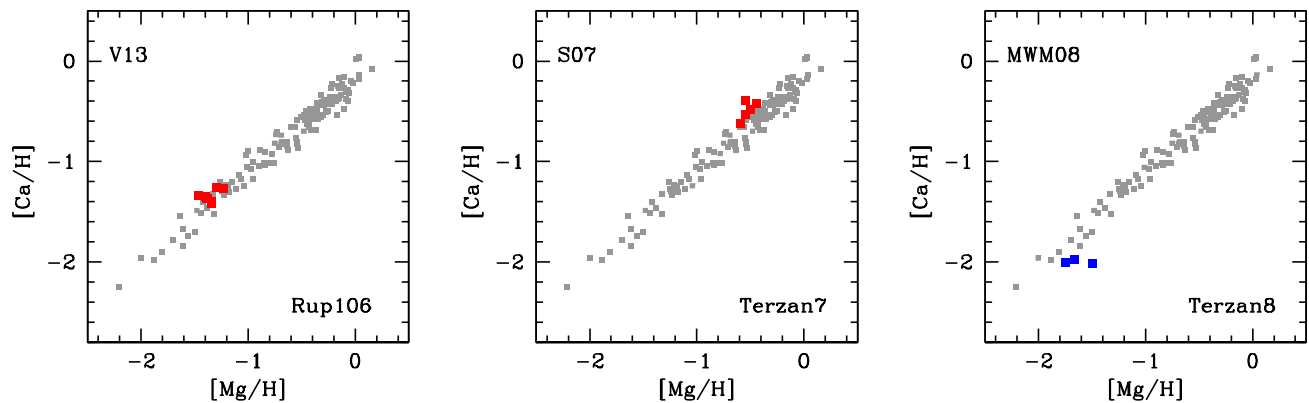} 
\caption{As in Fig.~\ref{f:A1} for the other GCs in the literature sample.}
\label{f:A10}
\end{figure*}

\clearpage

\section{View from Sc}

In this appendix we show the visual catalogue of the data for the Sc-Mg
distribution observed in all GCs examined in the present work. In Fig.~B1 we
show the clusters in the golden sample, and in Figs.~B2-B8 the same for the 
literature sample.

\begin{figure*}[h]
\centering
\includegraphics[scale=0.90]{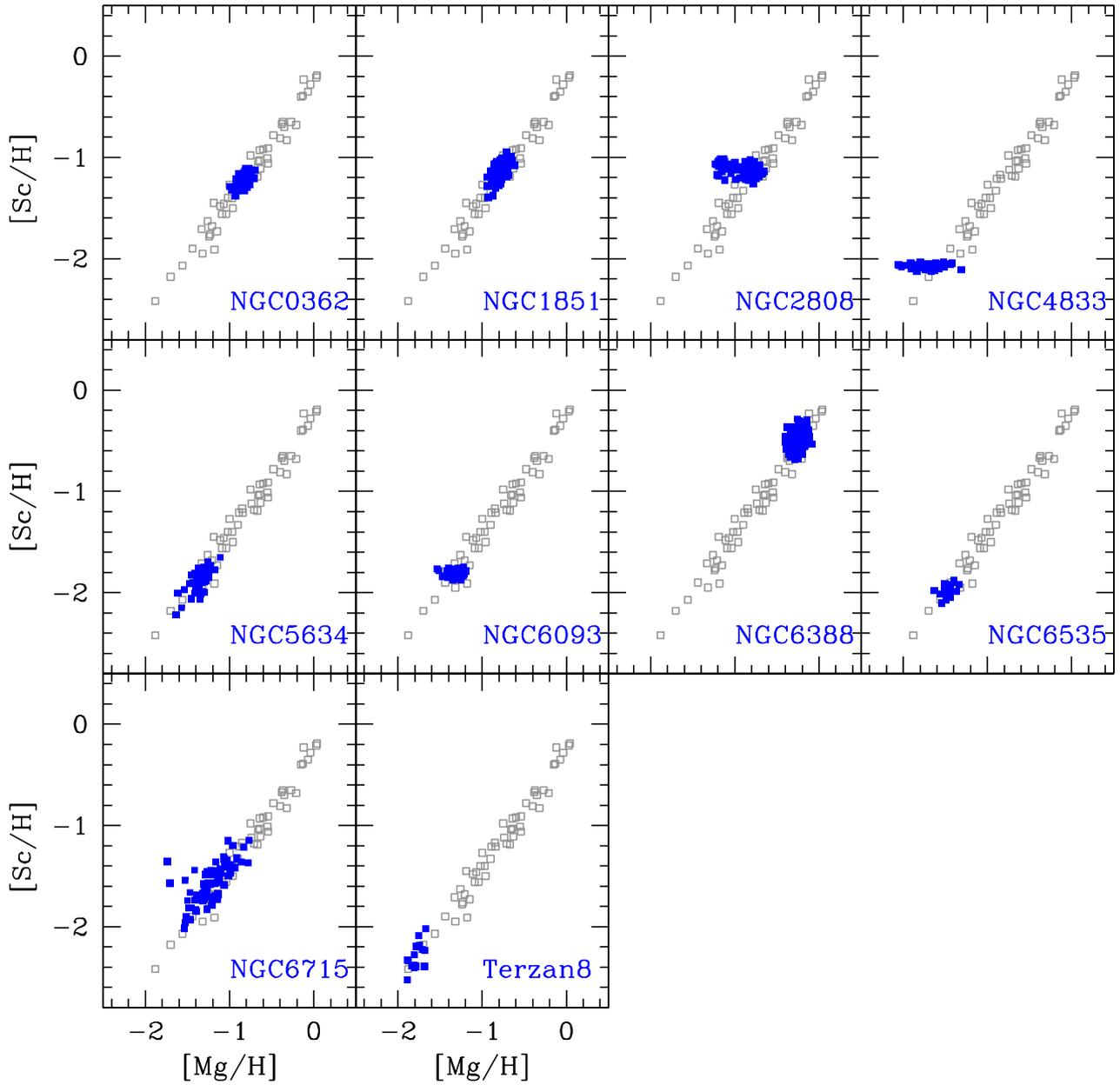}
\caption{Observed distributions [Sc/H] as a function of [Mg/H] ratios for  the
ten GCs in our golden sample, superimposed on field stars in Gratton et al.
(2003).}
\label{f:B1}
\end{figure*}

\begin{figure*}[h]
\centering
\includegraphics[scale=0.90]{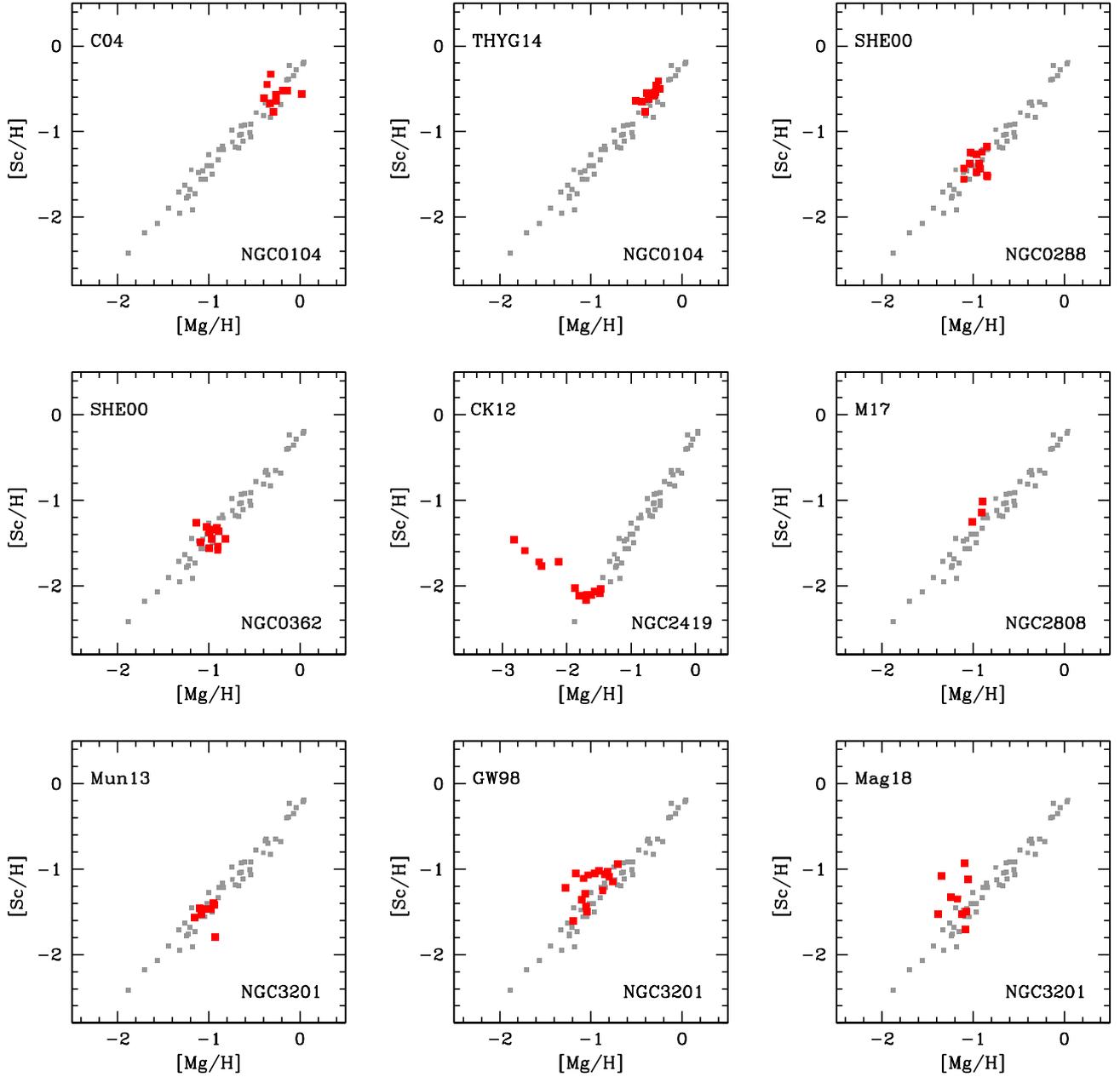}
\caption{Observed distributions [Sc/H] as a function of [Mg/H] ratios for  the
first nine GCs in the literature sample, superimposed on field stars in Gratton
et al. (2003). In each panel we list the alphanumeric code of Tab~\ref{t:tab1} to
identify the corresponding study. GC stars are shown in blue whenever the
sample includes fewer than five objects.}
\label{f:B2}
\end{figure*}

\begin{figure*}[h]
\centering
\includegraphics[scale=0.90]{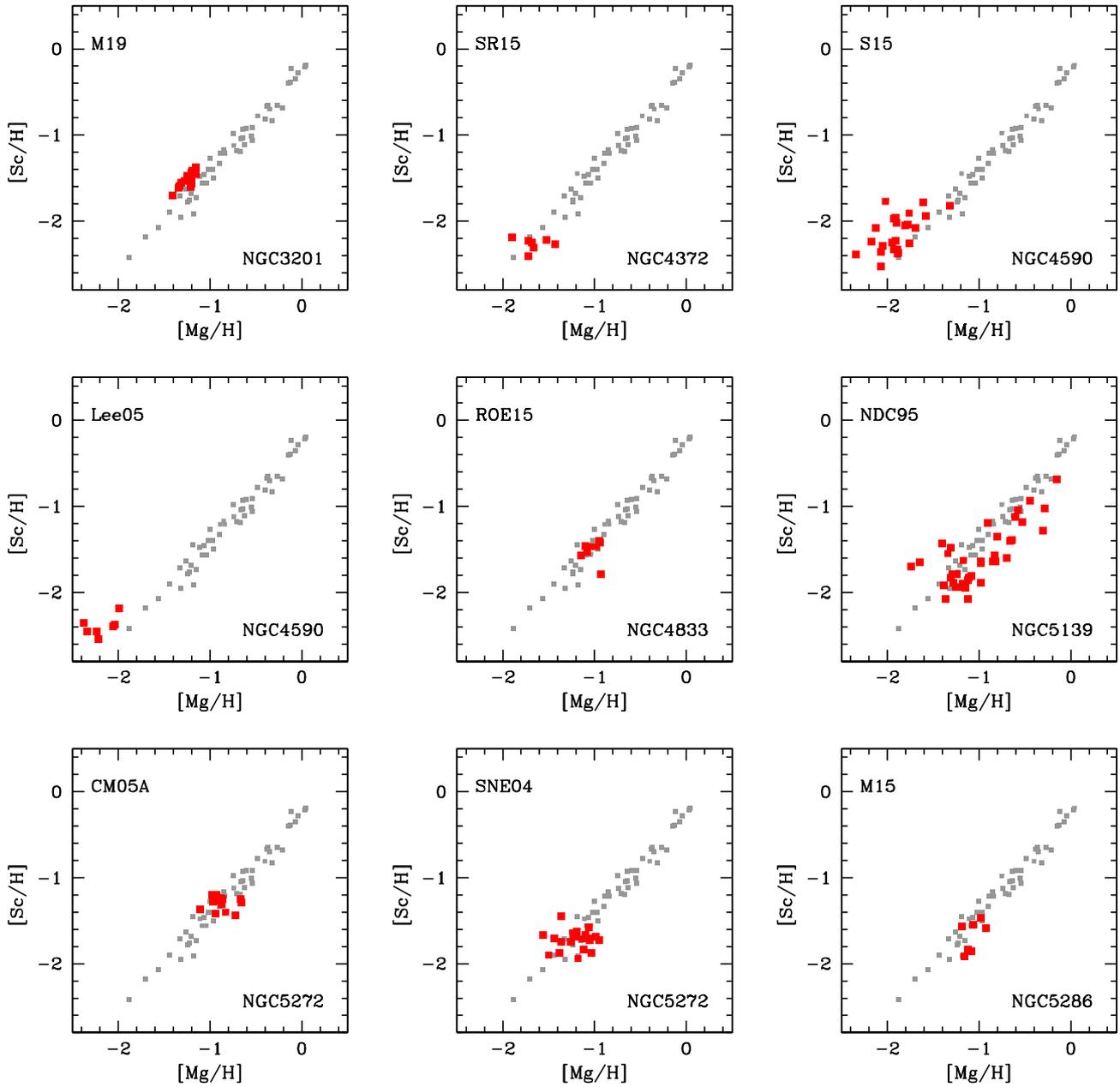}
\caption{As in Fig.~\ref{f:B2} for the other nine GCs.}
\label{f:B3}
\end{figure*}

\begin{figure*}[h]
\centering
\includegraphics[scale=0.90]{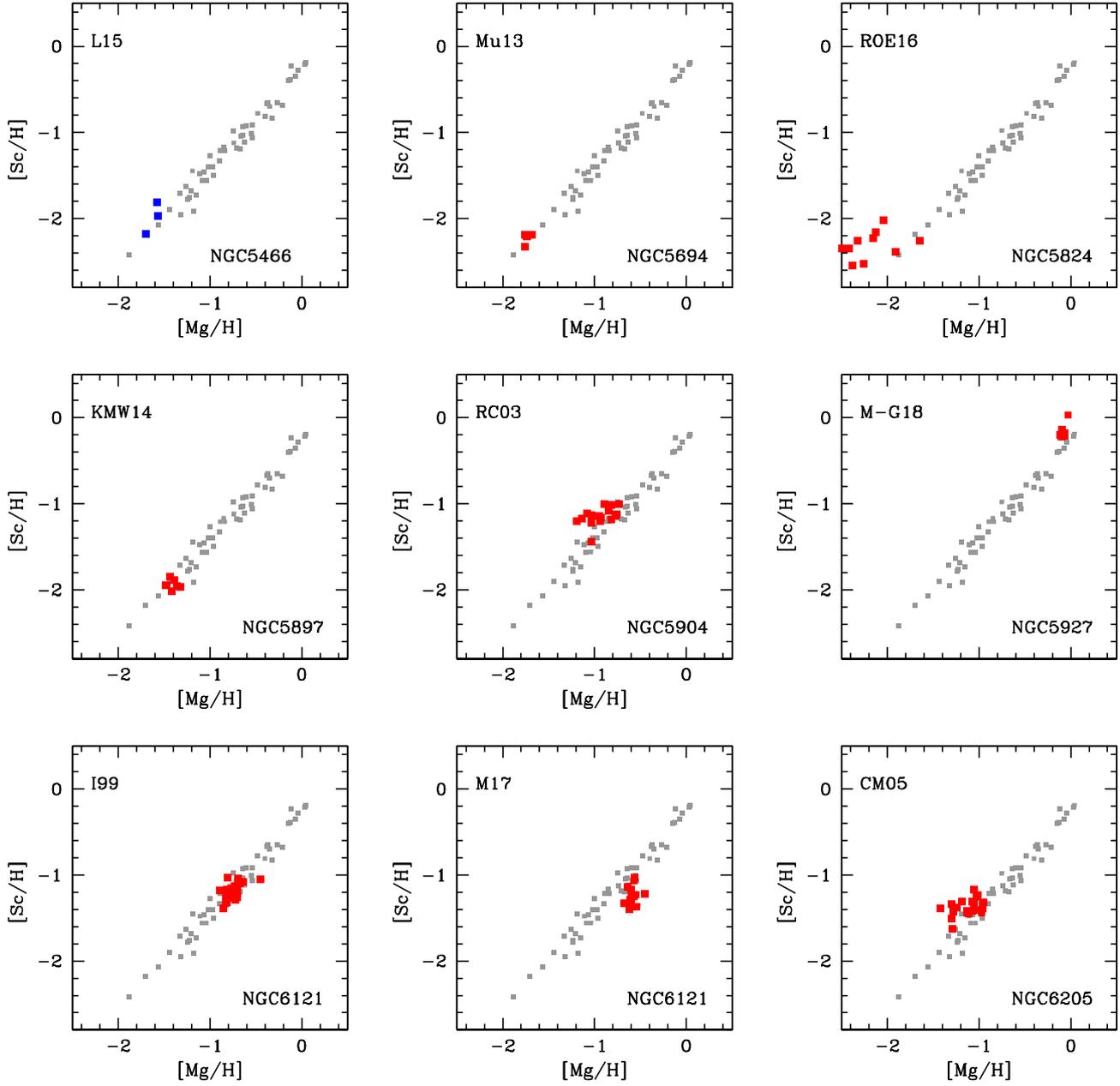}
\caption{As in Fig.~\ref{f:B2} for the other nine GCs.}
\label{f:B4}
\end{figure*}

\begin{figure*}[h]
\centering
\includegraphics[scale=0.90]{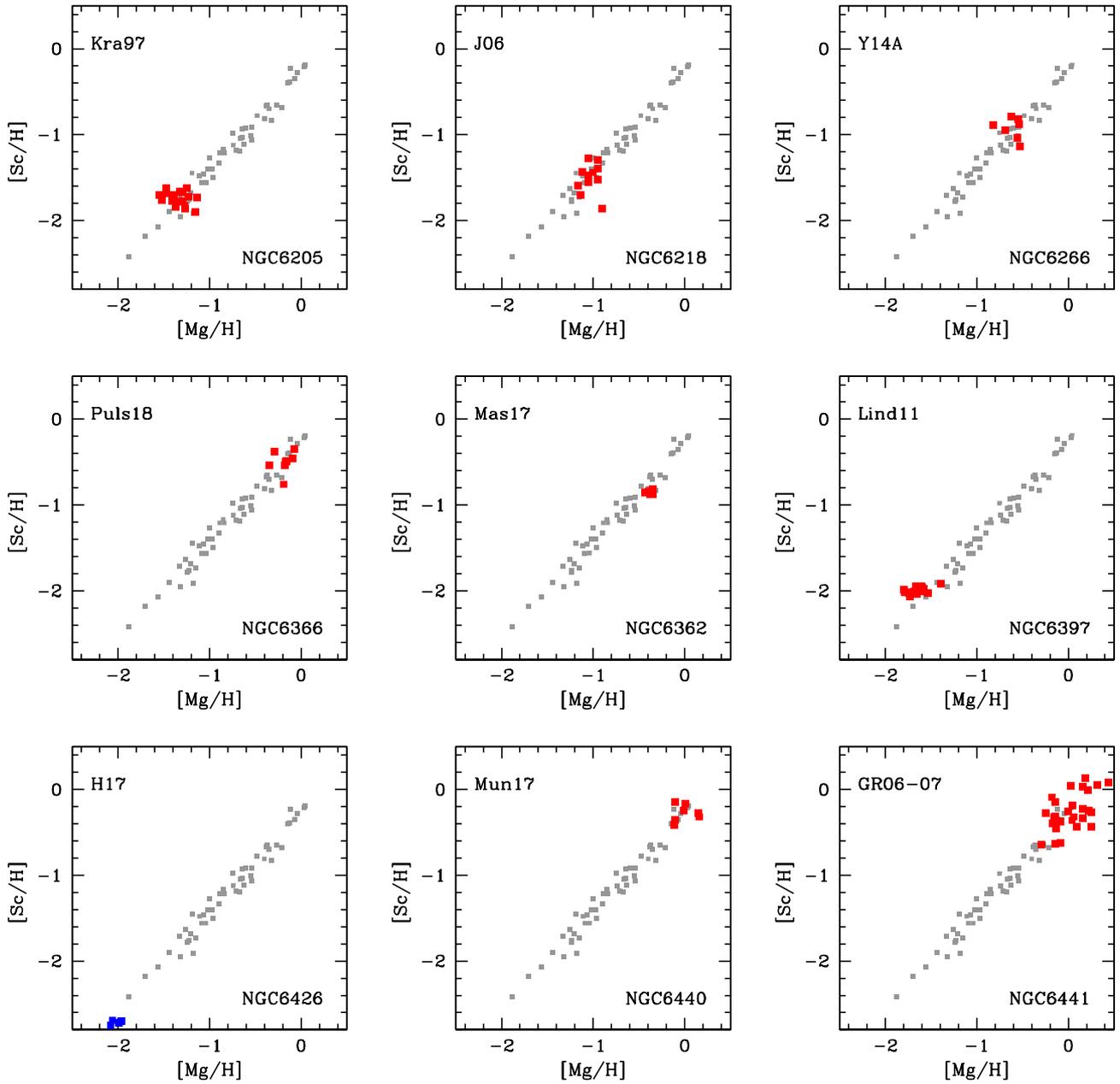}
\caption{As in Fig.~\ref{f:B2} for the other nine GCs.}
\label{f:B5}
\end{figure*}

\begin{figure*}[h]
\centering
\includegraphics[scale=0.90]{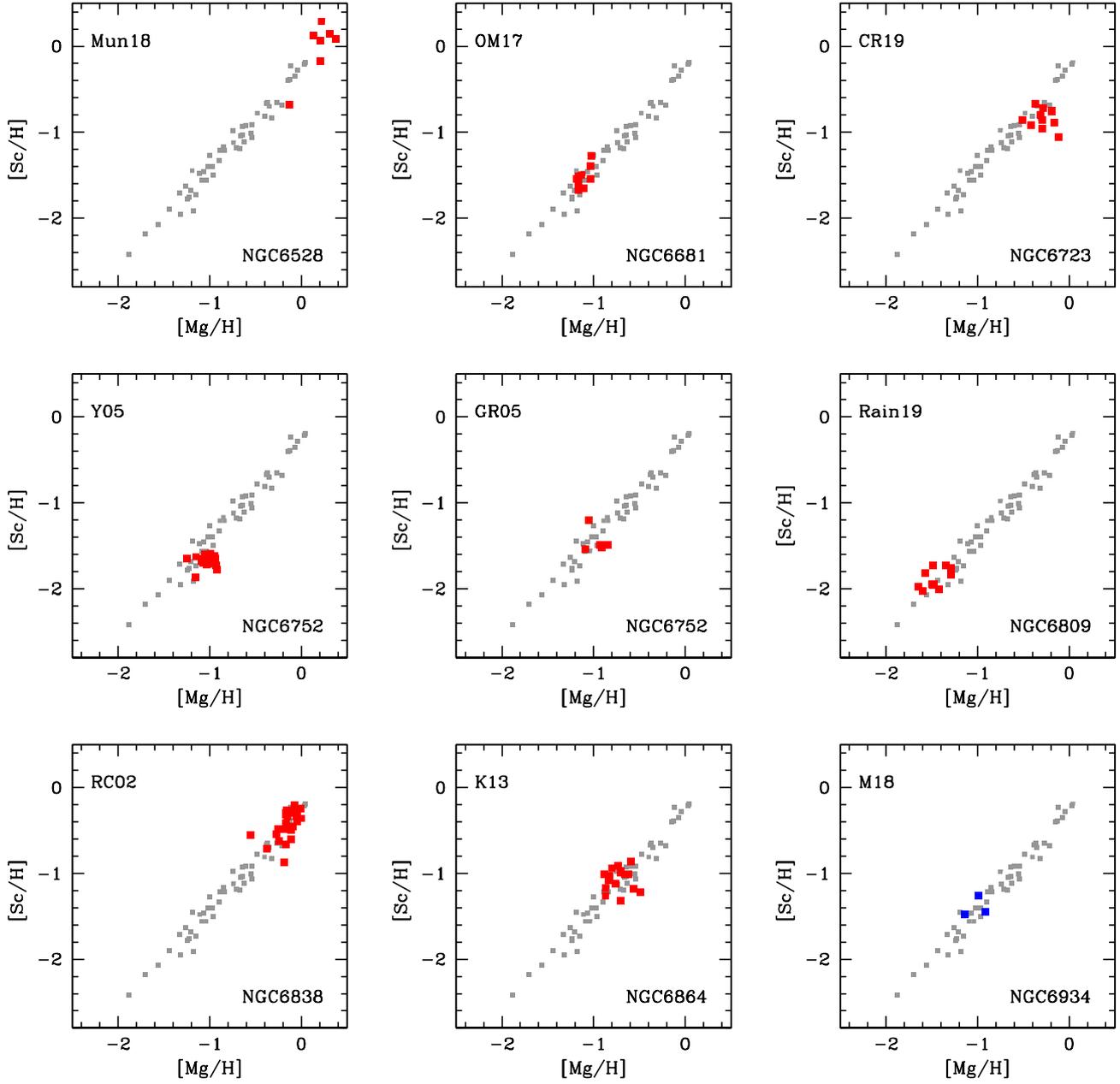}
\caption{As in Fig.~\ref{f:B2} for the other nine GCs.}
\label{f:B6}
\end{figure*}

\begin{figure*}[h]
\centering
\includegraphics[scale=0.90]{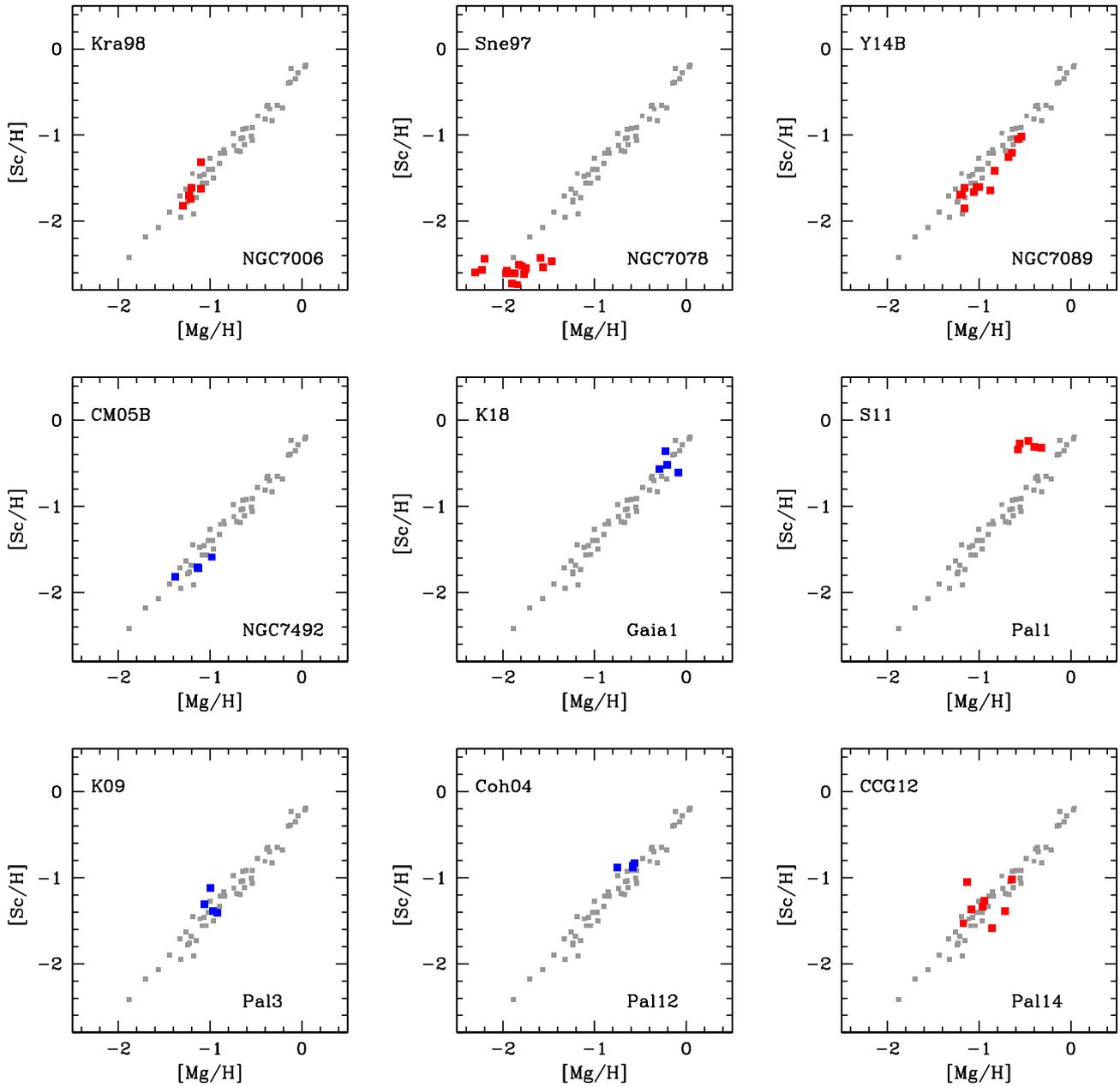}
\caption{As in Fig.~\ref{f:B2} for the other nine GCs.}
\label{f:B7}
\end{figure*}

\begin{figure*}[h]
\centering
\includegraphics[scale=0.90]{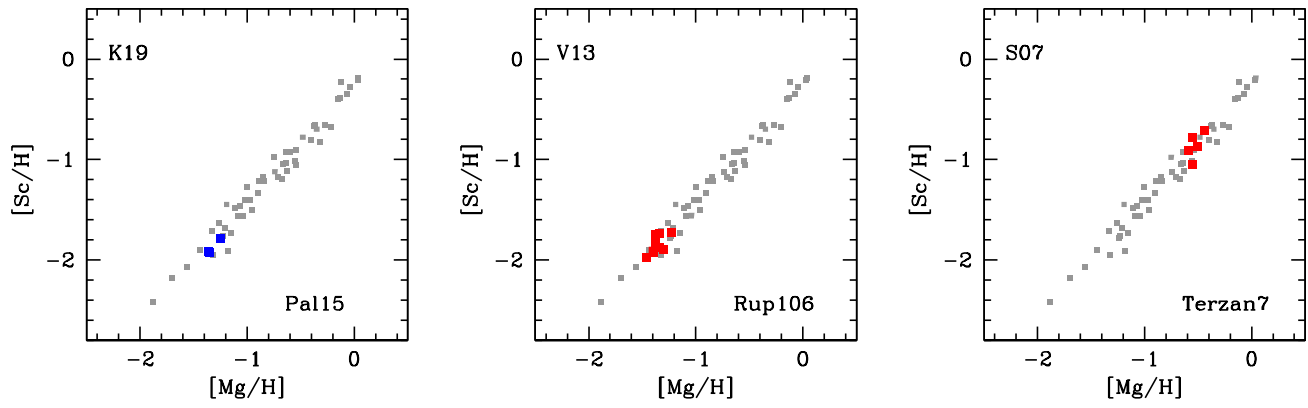}
\caption{As in Fig.~\ref{f:B2} for the last three GCs in the literature sample.}
\label{f:B8}
\end{figure*}

\clearpage

\section{Observed distributions from APOGEE}

In this appendix we show the catalogue of the observed Ca-Mg distribution for
the GCs in the APOGEE sample (Figs.~C1 and C2).

\begin{figure*}[h]
\centering
\includegraphics[scale=0.90]{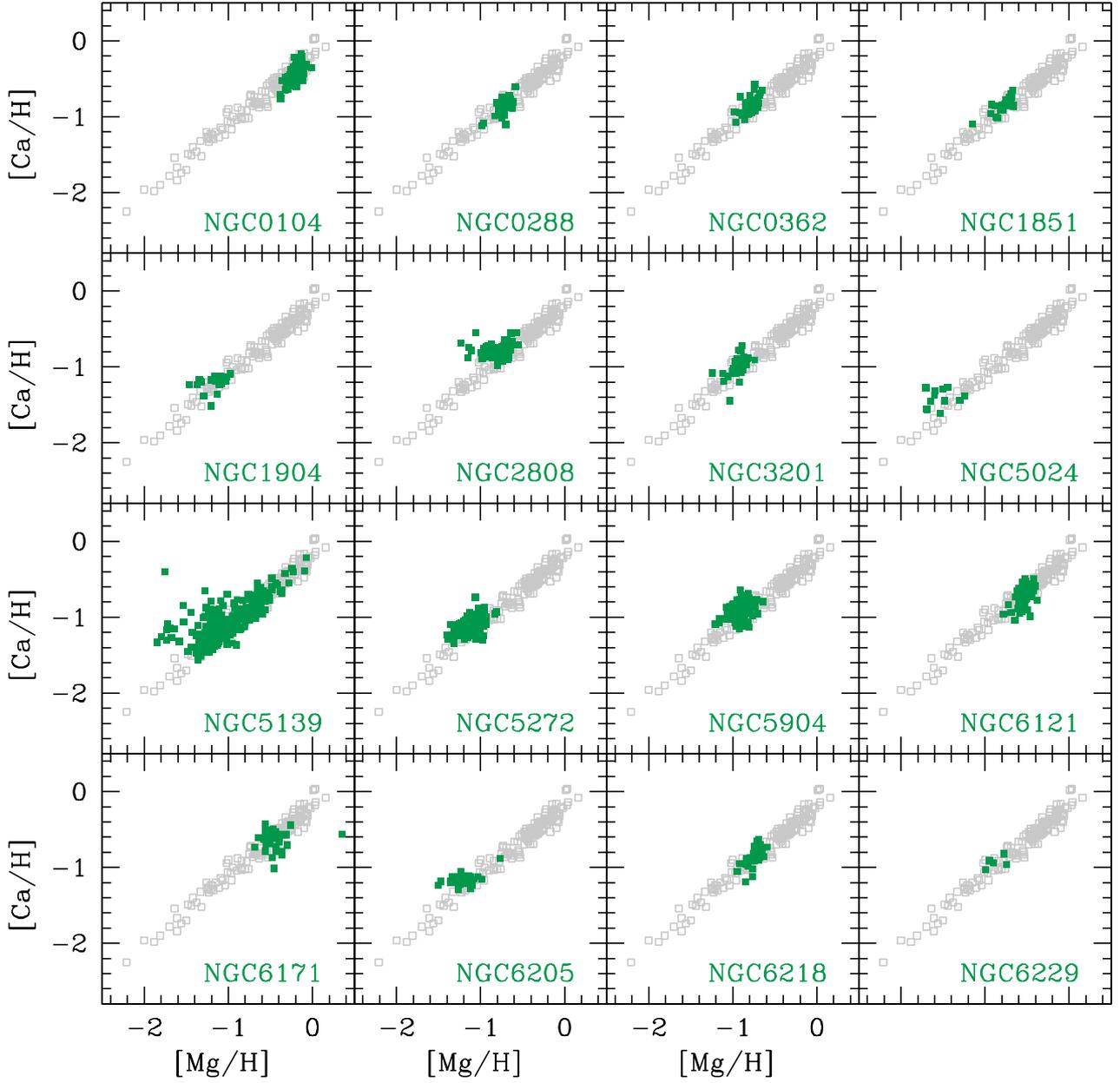} 
\caption{Observed [Ca/H] ratios as a function of [Mg/H] ratios for the  GCs in
the APOGEE-2 sample (M\'esz\'aros et al. (2020) compared to the field stars
in Gratton et al.  (2003).}
\label{f:apo1}
\end{figure*}

\begin{figure*}[h]
\centering
\includegraphics[scale=0.90]{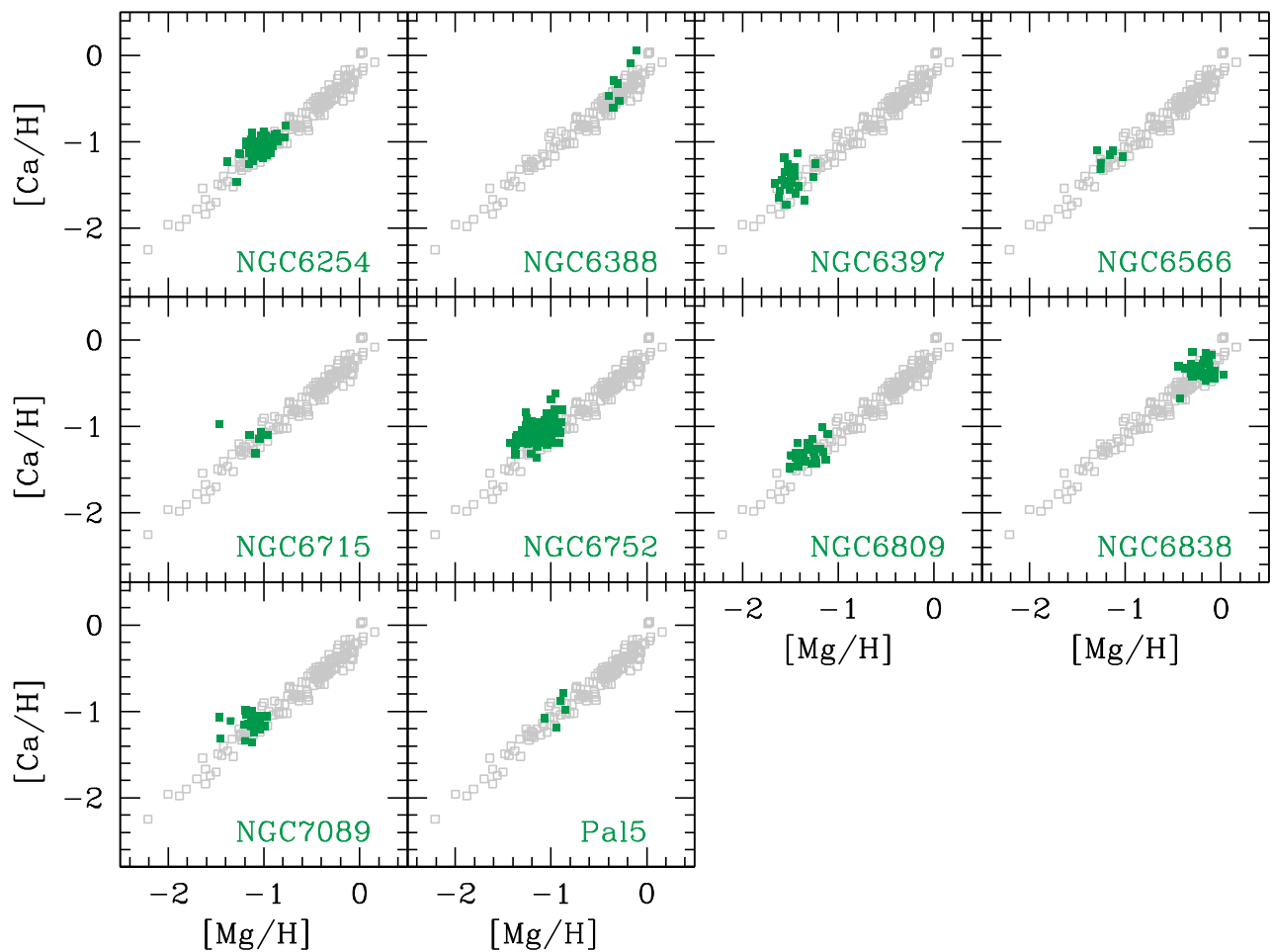} 
\caption{As in Fig.~C.1 for the other ten GCs in M\'esz\'aros et al. 
(2020).}
\label{f:apo2}
\end{figure*}

\end{appendix}

\end{document}